\documentclass[aps,prl,twocolumn,superscriptaddress]{revtex4-1}

\usepackage{amsmath}

\usepackage[pdftex]{graphicx}
\usepackage[thinspace]{SIunits}

\usepackage{tikz}

\newcommand*{\DashedArrow}[1][]{\mathbin{\tikz [baseline=-0.25ex,-latex, dashed,#1] \draw [#1] (0pt,0.5ex) -- (1.3em,0.5ex);}}%

\providecommand \BibitemShut  [1]{\csname bibitem#1\endcsname}%

\begin{document}


\title{Controlled tunneling induced dephasing of Rabi rotations for ultra-high fidelity hole spin initialization}
\author{P.-L. Ardelt}
 \affiliation{Walter Schottky Institut and Physik-Department, Technische Universit\"at M\"unchen, Am Coulombwall 4, 85748 Garching, Germany \\}
\author{T. Simmet}
 \affiliation{Walter Schottky Institut and Physik-Department, Technische Universit\"at M\"unchen, Am Coulombwall 4, 85748 Garching, Germany \\}
 \author{K. M\"uller}
 \affiliation{E. L. Ginzton Laboratory, Stanford University, Stanford, California 94305, USA\\}
\author{C.Dory}
 \affiliation{E. L. Ginzton Laboratory, Stanford University, Stanford, California 94305, USA\\}
 \affiliation{Walter Schottky Institut and Physik-Department, Technische Universit\"at M\"unchen, Am Coulombwall 4, 85748 Garching, Germany \\}
 \author{K.A. Fischer}
 \affiliation{E. L. Ginzton Laboratory, Stanford University, Stanford, California 94305, USA\\}
 \author{A. Bechtold}
 \affiliation{Walter Schottky Institut and Physik-Department, Technische Universit\"at M\"unchen, Am Coulombwall 4, 85748 Garching, Germany \\}
\author{A. Kleinkauf}
 \affiliation{Walter Schottky Institut and Physik-Department, Technische Universit\"at M\"unchen, Am Coulombwall 4, 85748 Garching, Germany \\}
 \affiliation{E. L. Ginzton Laboratory, Stanford University, Stanford, California 94305, USA\\}
\author{H. Riedl}
 \affiliation{Walter Schottky Institut and Physik-Department, Technische Universit\"at M\"unchen, Am Coulombwall 4, 85748 Garching, Germany \\}
\author{J.J. Finley}
 \affiliation{Walter Schottky Institut and Physik-Department, Technische Universit\"at M\"unchen, Am Coulombwall 4, 85748 Garching, Germany \\}
\email{finley@wsi.tum.de}

\date{\today}

\begin{abstract}

We report the sub-picosecond initialization of a single heavy hole spin in a self-assembled quantum dot with $>98.5 \%$ fidelity and \textit{without} external magnetic field. Using an optically adressable charge and spin storage device we tailor the relative electron and hole tunneling escape timescales from the dot and simultaneously achieve high-fidelity initialization, long hole storage times and high efficiency readout via a photocurrent signal. We measure electric field-dependent Rabi oscillations of the neutral and charged exciton transitions in the ultrafast tunneling regime and demonstrate that tunneling induced dephasing (TID) of excitonic Rabi rotations is the major source for the intensity damping of Rabi oscillations in the low Rabi frequency, low temperature regime. Our results are in very good quantitative agreement with quantum-optical simulations revealing that TID can be used to precisely measure tunneling escape times and extract changes in the Coulomb binding energies for different charge configurations of the quantum dot. Finally, we demonstrate that for sub-picosecond electron tunneling escape TID of a coherently driven exciton transition facilitates ultrafast hole spin initialization with near-unity fidelity. 

\end{abstract}


\maketitle

\section{Introduction}

The spin degree of freedom of charge carriers in semiconductor nanostructures is promising for the realization of quantum information technologies \cite{loss1998, kane1998}. In particular, spins locally trapped in optically active semiconductor quantum dots (QDs) have recently attracted strong interest, since their efficient coupling to light enables ultrafast spin control \cite{muller2013, press2008, press2010, DeGreve2011}, single shot read out of spin states \cite{vamivakas2010, delteil2014} and, more recently, the entanglement of stationary spins and photons \cite{ yamamoto2012, gao2012, steel2013} as the first building block for the realization of a quantum repeater \cite{mcmahon2015}. For all associated quantum protocols the high fidelity initialization of single spin states on ultra short timescales \cite{ muller2012, atature2006, gerardot2008} and the reliable storage of the spin state is crucial \cite{ kroutvar2004, heiss2009}. The initialization of single spins can either be performed by spin-pumping of a charged QD \cite{atature2006quantum} or by tunneling ionization of photo-generated excitons \cite{ ramsay2008fast, godden2012, godden2012fast, muller2012, mar2014ultrafast}. While spin pumping is very convenient since it requires only a single continuous wave laser, it is rather slow with reported initialization fidelities of $>90\%$ after $\sim 1ns$. On the other hand, tunneling ionization of photo-generated excitons without magnetic fields has been achieved with $>96\%$ preparation fidelities over picosecond timescales by exploiting ultrafast intra-molecular tunneling in quantum dot molecules \cite{muller2012} or shallow QDs \cite{mar2014ultrafast}. In both cases, the high fidelity is only achieved for specific experimental conditions that limit the hole lifetime. Moreover, the experiments presented in \cite{mar2014ultrafast} were performed with continuous wave excitation where the dynamics of the system can only be inferred with indirect techniques such as linewidth broadening \cite{oulton2002manipulation} due to the incoherent nature of the continuous wave - quantum dot interaction.

Here, we demonstrate the spin initialization of a single heavy hole spin with a preparation fidelity $> 98.5 \%$ on subpicosecond timescales \textit{without} applying a magnetic field. The heavy hole spin is initialized by partial tunneling ionization of excitons prepared using ultrafast optical methods in an InGaAs QD embedded in a Schottky diode structure \cite{godden2010, muller2012}. In order to maximize the spin initialisation fidelity, (i) we engineer the bandstructure to tailor the electron and hole tunnelling times from the QD by optimizing the Al content in an AlGaAs blocking barrier adjacent to the QD and (ii) we operate the Schottky diode structure in the ultrafast tunneling regime, where electron tunneling occurs on timescales similiar to the pulsed coherent excitation of an exciton. Accordingly, we experimentally record and theoretically model an intensity damping of excitonic Rabi oscillations due to tunneling induced dephasing (TID). By investigating electron tunnelling from a charged exciton state, we demonstrate that TID can \textit{vice versa} be used to determine electron tunneling times on timescales difficult to access by conventional pump-probe spectroscopy. Finally, we show, that in the TID-regime a single heavy hole spin can be initialized with fidelities exceeding $>98.5 \%$, overcoming prior limitations of the spin initialization fidelity due to the fine structure precession of the electron-hole pairs prior to ionization and making the initialization insensitive to errors in the excitation pulse intensity. 

\section{Experimental methods}

\begin{figure}[t]
\includegraphics[width=1\columnwidth]{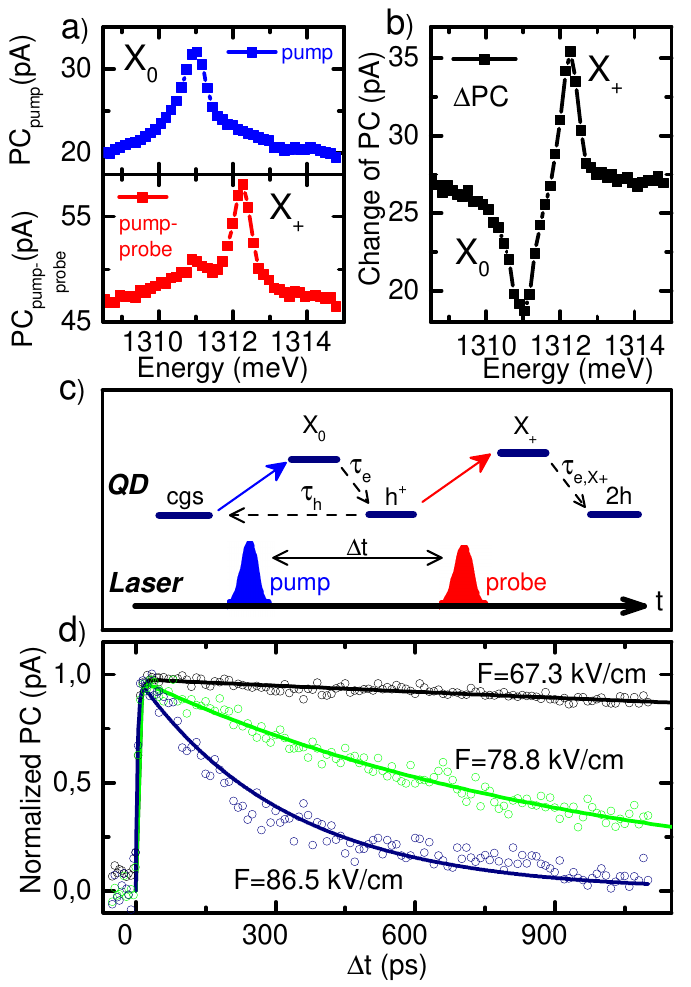}
\caption{\label{fig:Figure1} (a) PC absorption spectra for one (blue) and two pulse (red) excitation probing absorption the neutral exciton transition $X_{0}$ and positive trion transition $X_{+}$ respectively. (b) Difference of PC spectra $\Delta PC = PC_{pump-probe}-PC_{pump}$ for one and two pulse excitation. (c) Schematic illustration of the charge carrier dynamics probed in (b): the second excitation pulse at a delay time $\Delta t$ probes the positive trion $X_{+}$ as the electron tunnels out of the QD faster than the hole $\tau_{e} << \tau_{h}$. (d) PC amplitude of the trion transition $X_{+}$ as a function of the probe-pulse delay $\Delta t$. With increasing field the tunneling rate of the hole increases. Fits using a rate equation model are presented as solid lines.}
\end{figure}

The sample consists of single layer of self assembled InGaAs QDs embedded in a n-i Schottky diode \cite{warburton2000}. The QDs are placed 125nm above a heavily n-doped back contact and covered with a further 125nm thick layer of GaAs. For hole storage samples, we integrate an additonal 20nm think $Al_{0.1}Ga_{0.9}As$ barrier 10nm above the QDs into the GaAs layer as schematically illustrated on the right most panel of Figure \ref{fig:Figure2}a and b. The Schottky diode is formed by a $3nm$ semi-transparent Ti layer covered with a shadow mask of Au with circular $1-2\mu m$ diameter apertures to ensure that we optically address only single QDs \cite{Krenner2005}. Most importantly, applying a bias between the Au top contact and the n-doped back contact facilitates control of the local electric field F in growth direction at the QDs and thus control the electron and hole tunneling times $\tau_{e}$ and $\tau_{hh}$ from the QDs. For $\tau_{e}$ shorter than the radiative recombination time $\tau_{rad} > \tau_{e}$ of the electron-hole pairs, the Schottky diode structure is operated in the photocurrent (PC) regime. There, resonant absorption of the excitation laser leads to a measurable change of current in the external measurement circuit \cite{zrenner2002}. 

Typical raw data obtained in such a PC experiment is presented in figure \ref{fig:Figure1}a. The figure shows the PC induced in the sample when exciting a single QD in the PC-regime with a train of 5ps laser pulses (pump). To achieve continuous tuning of the laser wavelength the pulses are derived from a 150fs broadband Ti:saphire laser source (repetition rate 79MHz) by using a 4f-pluse shaping geometry \cite{muller2013, muller2013probing, godden2012coherent}. A clear peak is observed at $E=1311.0meV$ arising from resonantly driving the neutral exciton transition $X_{0}$ in the QD. In order to investigate the dynamics of charge carriers excited by the pump pulse train, we add a second train of laser pulses delayed by a time $\Delta t$ as schematically illustrated in figure \ref{fig:Figure1}c. A typical PC-absorption spectrum for two-pulse excitation with the first pulse fixed to the $cgs \rightarrow X_{0}$ transition and the second pulse (probe), delayed by $\Delta t =25ps$ and tunable in energy, is presented in figure \ref{fig:Figure1}a in red. Comparing the one and two pulse PC absorption spectra in figure \ref{fig:Figure1}a, shows that the amplitude of the neutral exciton transition $X_{0}$ is strongly suppressed, while an additional peak emerges at $E=1312.3 meV$, detuned by $1.3 meV$ from the neutral exciton transition. This additional peak results from the single hole $h+$ to positively charged trion transition $X_{+}$ as will be discussed below. The change of the absorption spectrum becomes even clearer by plotting the change of PC, namely $\Delta PC = PC_{pump-probe}-PC_{pump}$ as presented in figure \ref{fig:Figure1}b. Dips correspond to PC absorption for excitation with a single laser pulse (pump) and peaks to conditional PC absorption for excitation with a second laser pulse (probe).

The conditional bleaching of the $X_{0}$ amplitude and emergence of the $X_{+}$ absorption by a second excitation pulse can be understood from the dynamics illustrated in Figure \ref{fig:Figure1}c: The population of the QD, initially in the crystal ground state $cgs$, is transferred to the neutral exciton $cgs \rightarrow X_{0}$ with a Rabi rotation around $\pi$ by the pump-pulse (blue arrow in Figure \ref{fig:Figure1}c) and decays by subsequent tunneling of the electron and the heavy hole. This occurs with characteristic timescales $\tau_{e}$ for the electron tunneling $X_{0} \DashedArrow[->,densely dashed    ]$ $h^{+}$ and $\tau_{h}$ for the hole tunneling $h^{+} \DashedArrow[->,densely dashed    ]$ $cgs$ back to the $cgs$ \cite{godden2012, muller2012phonon}, where dashed arrows indicate transitions due to tunneling in contrast to full arrows indicating optically driven transitions. Primarly due to the lighter electron effective mass of $m^{*}_{e}= 0.05 m_{0}$ compared to the hole $m^{*}_{hh}= 0.34 m_{0}$ ($m_{0}$ denotes the free electron mass) \cite{goldberg1999handbook}, the electron tunneling time is typically shorter than the heavy hole hole tunneling time $\tau_{e} < \tau_{h}$ \cite{muller2012}. When the second pulse arrives with a time delay $\tau_{e} < \Delta t < \tau_{h}$, the neutral exciton transition $cgs \rightarrow X_{0}$ can not be driven as the QD is occupied with a single hole $h^{+}$. However, the probe pulse can excite the single hole to positive trion transition $h^{+} \rightarrow X_{+}$, that is positively detuned by an energy $\Delta E = 1.3meV$ from the neutral exciton $X_{0}$ (c.f. red arrow in figure \ref{fig:Figure1}c). The difference in energy results from the Coulomb and exchange interaction with the additional hole $h^{+}$. Thus, monitoring the amplitude of the $X_{+}$ transition as a function of the probe laser delay $\Delta t$ allows us to extract: (i) the electron tunnelling time $\tau_{e}$ from the rise time of the $X_{+}$ amplitude as electron tunneling from the neutral exciton state $X_{0}$ enables the $h^{+} \rightarrow X_{+}$ transition and (ii) the heavy hole tunnelling time $\tau_{h}$ from the decay time of the $X_{+}$ amplitude. 

In figure \ref{fig:Figure1}d, we present the time evolution of the amplitude of the trion transition $h^{+} \rightarrow X_{+}$ as a function of the time delay $\Delta t$ between the pump and probe pulse at three different electric fields $F$. The amplitude of the $X_{+}$ transition in figure \ref{fig:Figure1}d rises immediately corresponding to a very fast electron tunneling time ($\tau_{e} < 5ps$). The decay time due to hole tunnelling is much longer and decreases from $\tau_{h}=9.8ns$ to $\tau_{h}=322ps$ when increasing the internal electric field across the Schottky diode from $F=67.3$ keV/cm up to $F=86.5$ kV/cm. Note however, that quantitative access to the electron tunneling times $\tau_{e}<\tau_{pulse}=5ps$ is impossible, since the time resolution of pump probe experiments is inherently limited by the pulse length $\tau_{pulse}$. In order to extract the tunneling times $\tau_{e}$ and $\tau_{h}$, we model the time evolution of the $X_{+}$ transition amplitude with a rate equation model incorporating the populations of neutral exciton $X_{0}$, the positively charged trion $X_{+}$ and the crystal ground state $cgs$ (see supplementary material) \cite{muller2012phonon}. 

\section{Optimized spin storage device for photocurrent read out}

\begin{figure}[t]
\includegraphics[width=1\columnwidth]{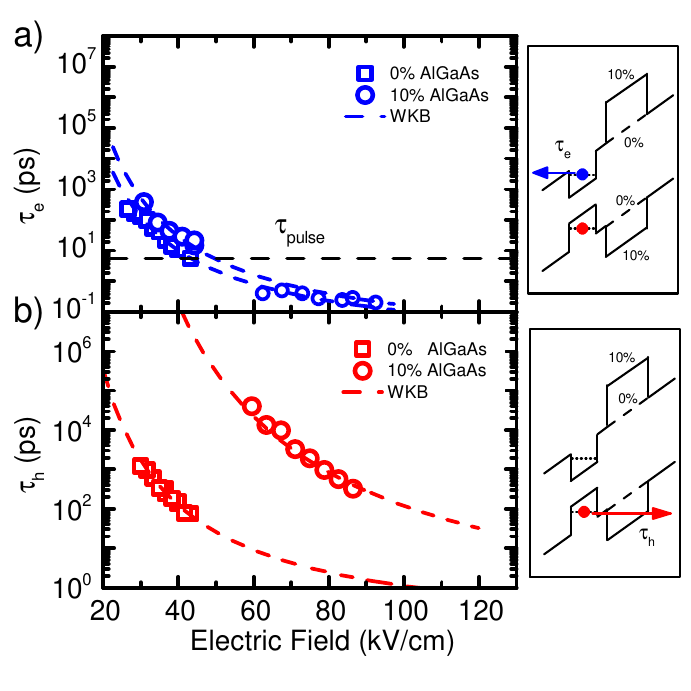}
\caption{\label{fig:Figure2} (Color online) Comparison of the field dependent tunneling times for two samples with and without the 10 $\%$ AlGaAs barrier. (a) Electric field dependent electron and (b) hole tunneling times. Fits to the data using the WKB model are plotted as dashed lines.}
\end{figure}

In order to use the Schottky diode structure as a spin storage device, independent engineering of the tunneling times $\tau_{e}$ and $\tau_{h}$ is crucial. Ideally, for a hole spin storage device, we aim for two properties: (i) An ultrafast electron tunneling time $\tau_{e}$, since the electron tunneling time $\tau_{e}$ determines the timescale on which a single hole spin can be initialized by tunneling ionization of the neutral exciton $X_{0}$ and (ii) a long hole storage time $\tau_{h}$ that facilitates use of the hole spin qubit. The individual tunneling times $\tau_{e}$ and $\tau_{h}$ can be tailored by incorporating an $Al_{x}Ga_{1-x}As$ barrier immediately adjacent to the QD layer \cite{heiss2009}. As schematically illustrated in figure \ref{fig:Figure2}a and b for a heavy hole tunneling barrier, the charge carrier tunneling out of the QD faces an additional barrier with a height that depends on the $Al$ concentration $x$ of the barrier allowing to design the asymmetry of the tunneling times a priori using WKB theory \cite{fry2000, villas2005}.

In figure \ref{fig:Figure2}a and \ref{fig:Figure2}b, we present the electron and hole tunneling times $\tau_{e}$ and $\tau_{h}$ for two devices: One n-i Schottky diode with $125nm$ GaAs below and above the QD layer respectively (results shown as squares) and a second diode, where a 20nm thick tunneling barrier with $10 \%$ Al concentration was inserted 10nm above the QD layer (results shown as circles). To be able to compare the tunneling times, the two QDs compared have similar transition energies of the neutral exciton $X_{0}$.

The electron tunneling times $\tau_{e}$ presented in figure \ref{fig:Figure2}a extracted from the rise time of the $X_{+}$ amplitude and cw-PC measurements of the linewidth of $X_{0}$ (see supplementary material), exhibit a similar electric field dependence, demonstrating that electron tunneling is not affected by barriers above the QDs. In contrast, the heavy hole tunneling time $\tau_{h}$ in figure \ref{fig:Figure2}b exhibits an approximately three orders of magnitude increase due to the $Al_{0.1}Ga_{0.9}As$ barrier. To quantitatively analyse the data we calculate the tunneling times using a WKB model \cite{fry2000, villas2005}:

\begin{equation}
\tau_{e (h)}(F)=\frac{2m_{e (h)}^{*}L^{2}}{\hbar\pi}exp\left[\frac{4}{3\hbar e F}\sqrt{2m_{e (h)}^{*}E_{i}^{3}}\right]
\label{eq:WKB}
\end{equation}

where $m_{e (h)}^{*}$ denotes the effective mass of electrons (heavy holes), $L$ the effective width of the QD potential in z-direction, $F$ the electric field and $E_{i}$ the height of the triangular barrier in the conduction (valence) band. Fits to the data are presented as dashed lines and produce good overall agreement with our experimental data. 

Modeling the effect of the $Al_{0.1}Ga_{0.9}As$ barrier on the hole tunneling time, we obtain an effective mass of $m_{h}^{*}=0.34 m_{0}$ for the heavy holes and an effective width for the QD potential of $L = 5.1nm$ \cite{muller2012phonon}, when increasing the height of the triangular barrier in the valence band from $E_{i, 0\%}=39.8meV$ to $E_{i,10\%}=73.0meV$. The increase of the triangular barrier height $\Delta E_{i} = 33.2 meV$ qualitatively agrees with the valence band offset $\Delta E_{v} = 40meV$ due the $10 \%$ Al-concentration in the barrier, while the effective width $L$ and effective mass are in good agreement with previously reported values for quantum dots with a similar material composition \cite{muller2012phonon}.

\section{Tunnelling induced dephasing of Rabi rotations}

\begin{figure}[t]
\includegraphics[width=1\columnwidth]{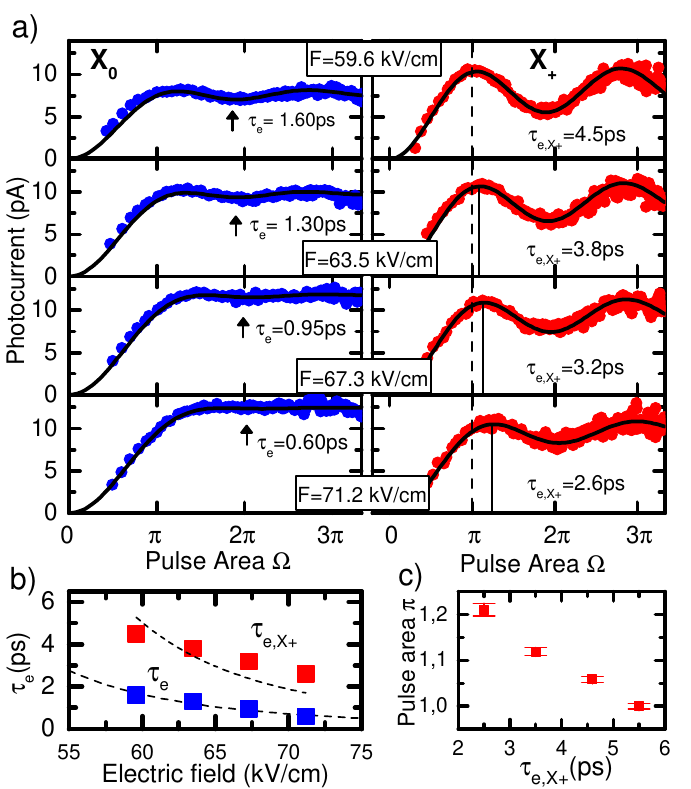}
\caption{\label{fig:Figure3} (a) Damped Rabi rotations of the neutral $X_{0}$ and charged exciton $X_{+}$. Increasing the electric field leads to faster electron tunneling $\tau_{e}$ and $\tau_{e,X_{+}}$ and stronger TID. (b) Electron tunneling times extracted from modelling the TID for the neutral $\tau_{e}$ and charged $\tau_{e,X_{+}}$ exciton. (c) Renormalization of the area of a $\pi$ pulse due to increased dephasing of the Rabi rotations.}
\end{figure}

To minimize the hole spin initialization time, we operate the Schottky diode in the electric field regime $F > 50 kV/cm$, where the electron tunnelling time $\tau_{e}$ corresponding to the ionization time of the neutral exciton $X_{0}$, is shorter than the pulse length of the excitation laser $\tau_{pulse}$. However, when the interaction time between QD and laser becomes comparable to the lifetime of the driven state (which is given by $\tau_{e}$), this can no longer be neglected as a source of dephasing \cite{ramsay2010damping, villas2005}. The blue data points in figure \ref{fig:Figure3}a show the photocurrent as a function of the pulse area $\Omega$ of the driving laser in resonance with the neutral exciton $X_{0}$ and electric fields in the range $F=59.6 kV/cm$ to $F=71.2 kV/cm$; an electric field region where the electron tunneling time $\tau_{e}$ is clearly shorter than the excitation laser pulse lengths of $\tau_{pulse} \cong 5 ps$. While for an electric field of $F=59.6 kV/cm$ in Figure \ref{fig:Figure3}a, a damped $2\pi$ Rabi rotation indicated by the black arrow can still be resolved \cite{ardelt, ramsay2010damping}, increasing the electric field to $71.2 kV/cm$ leads to a complete dephasing of the Rabi rotations. Importantly, the current does not converge towards a value that is lower than the maxima of the oscillations but instead converges towards the maximum given by the repetition rate of the laser. Also note that the photocurrent readout does not correspond to the probability of the exciton occupation after the interaction with this pulse but rather of the probability of not being in the ground state (i.e. in the exciton or single hole state).

We continue to show that the damping of the observed oscillations, as well as convergence to the maximum value results exclusively from the fast electron tunneling. Typically the major source of dephasing for excitonic Rabi rotations is coupling to LA-phonons leading to an amplitude damping of the Rabi oscillations in the frequency domain \cite{ramsay2010damping, ardelt}. However, as we operate the QD at a low temperature of $T=4.2K$ and for pulse areas up to $3\pi$, corresponding to the low Rabi frequency limit \cite{ramsay2010phonon}, the according excitation induced dephasing time due to LA-phonon coupling is weak. The corresponding dephasing time can be calculated to be smaller than $\Gamma_{2,phonons}^{-1}= K_{2}(T=4.2K) \Omega_{3\pi}=35.3 ps$  for the peak Rabi frequency $\Omega = 3\pi$ of a $5ps$ long excitation pulse and a material dephasing time constant $K_{2}(T=4.2K) \cong 0.1 ps$ for InGaAs QDs at $T=4.2 K$ \cite{ramsay2010damping}. Comparing this excitation induced dephasing time $\Gamma_{2,phonons}^{-1}$ of $35.3ps$ to the electron tunneling times of $\tau_{e}<5ps$ clearly identifies TID as the major source of dephasing for the neutral exciton Rabi rotations presented in figure \ref{fig:Figure3}a. To model the TID quantitatively, we performed quantum optical simulations using the Quantum Optical Toolbox in Python (Qutip) \cite{qutip}. We utilized the Lindblad form of the quantum optical master equation to simulate Rabi oscillations of the charge carrier dynamics described in figure \ref{fig:Figure1}c. TID was taken as a phenomenological dephasing rate and was extracted at different field strength through fits to the experimental data in figure \ref{fig:Figure3}a. Due to the extremely long hole tunneling time and high fidelity of initialization, each simulation was reduced to a three level subsystem. The result of these simulations is presented as  black lines on top of the data in Figure \ref{fig:Figure3}a and produces excellent agreement. The tunneling times extracted from the simulations are presented in Figure \ref{fig:Figure3}b as blue datapoints and range from $\tau_{e}=1.60ps$ at $F=59.6 kV/cm$ to $\tau_{e}=0.60ps$ at $F=71.2 kV/cm$. The WKB fit from Figure \ref{fig:Figure2}a is reproduced in this figure as well and produces very good quantative agreement. 

We proceed by utilizing the TID of the Rabi rotations to precisely measure the electron tunneling times $\tau_{e,X_{+}}$ from the positively charged trion state $X_{+}$ (schematically shown in Figure \ref{fig:Figure1}c). Therefore, we present the Rabi rotations of the positively charged exciton state $X_{+}$ as a function of pulse area of the probe pulse in red in Figure \ref{fig:Figure3}a. Similar to the neutral exciton $X_{0}$, the intensity of the Rabi oscillations is progressively damped when increasing the electric field from $F=59.6 kV/cm$ to $F=72.2 kV/cm$. However, comparing the damping of the neutral exciton $X_{0}$ and charged state $X_{+}$ Rabi oscillations in Figure \ref{fig:Figure3}a, we clearly observe a stronger TID for the neutral exciton $X_{0}$. This indicates that electron tunneling from $X_{+}$ occurs significantly more slowly than from $X_{0}$. In order to extract these electron tunneling times $\tau_{e,X_{+}}$ from the charged exciton state $X_{+}$, we fit the data with the model described above. The fits are presented as black lines on top of the experimental data in Figure \ref{fig:Figure3}a and produce excellent agreement, as for the neutral exciton $X_{0}$. 

The extracted tunneling times from the charged exciton $X_{+}$ state, are presented in Figure \ref{fig:Figure3}b as red data points. Comparing them to the values obtained for electron tunnelling from $X_{0}$ (Figure \ref{fig:Figure3}b – blue) shows that for the same electric field tunnelling from $X_{+}$ occurs significantly more slowly than from $X_{0}$. The increased electron tunnelling times $\tau_{e,X_{+}}$ from the charged exciton state $X_{+}$ can be understood in the following way: For tunnelling from the charged state  $X_{+} \DashedArrow[->,densely dashed    ]$ $2h$ illustrated in Figure \ref{fig:Figure1}b an additional Coulomb attraction from the second heavy hole in the QD has to be overcome by the electron. We calculate the additional Coulomb attraction by fitting the electron tunneling times $\tau_{e,X_{+}}$ presented in Figure \ref{fig:Figure3}b  with a WKB fit where the barrier height $E_{i}$, corresponding to an additional inonization energy here, is varied while all other parameters are kept from the fit presented above (Figure \ref{fig:Figure2}a). This results in an additional ionization energy $\Delta E_{i}=11.2 meV$ which is in agreement with prior calculations of the additional exciton binding energy of second heavy hole $2h$ present in the QD including spatial rearrangement of the electron and hole wave functions \cite{finley2004quantum}. Note that due to the spatial rearrangement of the strongly localized hole wavefunctions in the QD, the energy reduction due the Coulomb attraction between $1e$ and the second heavy hole $2h$ exceeds the energy increase due to the Coulomb repulsion between the two heavy holes $2h$ leading to the observed increase in ionization energy $\Delta E_{i}$.

Closer inspection of figure \ref{fig:Figure3}a also reveals that the maxima of the oscillations shift to higher powers for larger electric fields. To analyse this in more detail, we present in Figure \ref{fig:Figure3}c the pulse area of a $\pi$ pulse for the $X_{+}$ transition as a function of the tunnelling time normalized to the pulse area of a $\pi$ pulse obtained at $F=59.6$ kV/cm. The reduction of Rabi frequency results from a renormalization due to dephasing, similar to renormalization resulting from strong dephasing due to LA-phonons at elevated temperatures \cite{ramsay2010phonon}.  

\section{Subpicosecond high fidelity hole spin initialisation}

\begin{figure}[t]
\includegraphics[width=1\columnwidth]{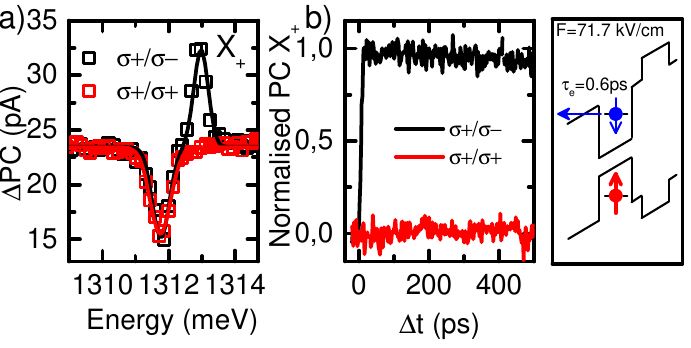}
\caption{\label{fig:Figure4} (a) Co- and cross-circularly polarized pump-probe measurements of the trion transition $X_{+}$. The complete suppression of the trion amplitude $X_{+}$ for a co-polarized excitation due to the Pauli blockade demonstrates high-fidelity spin initialization. (b) Time evolution of the $X_{+}$ amplitudes from (a) and a schematic illustration of the hole spin initialization due to electron tunneling ionization. The rise time corresponds to the spin initialization time. The suppression of the $h^{+} \rightarrow X_{+}$ transition amplitude driven with $\sigma_{+} / \sigma_{+}$ polarized light indicates high fidelity spin initialization and storage. }\end{figure}

Finally, we demonstrate high fidelity hole spin initialization in the TID regime $\tau_{e}<\tau_{pulse}$. Thereby, we excite the QD with $\sigma_{+}$ polarized excitation pulse to create a neutral exciton $X_{0}$ with a spin configuration $\downarrow \Uparrow$, as illustrated schematically in figure \ref{fig:Figure4}. The exciton spin state coherently precesses due the e-h exchange interaction with a period of $78ps$ between $\downarrow \Uparrow$ and $\uparrow \Downarrow$ (see supplementary material) \cite{muller2013}. However, if the exciton $X_{0}$ is ionized by electron tunnelling on timescales of $\tau_{e}<<78ps$, the remaining hole in the QD has a very well defined spin $\Uparrow$ projection parallel to the initially generated exciton $\downarrow \Uparrow$, defined by the optical axis and circular polarisation of the excitation source. The fidelity of the hole spin $\Uparrow$ initalization thus depends on $\tau_{e}$, since any precession of the spin wavefunction prior to the tunneling ionization will result in a statistical mixture of $\Uparrow$ and $\Downarrow$ for the hole spin \cite{godden2012, godden2010, muller2012}. Note here, that the TID makes the spin preparation insensitive to errors in the intensity of the pulse exciting the subsequently ionized exciton as can be seen in Figure \ref{fig:Figure3}a.  

To investigate the spin initialization fidelity in the regime $\tau_{pulse}>\tau_{e}$, we performed cross- and co-circulary polarized pump-probe measurements on the positively charged trion transition $X_{+}$. A typical spectrum recorded at an electric field $F=71.7 kV/cm$ is shown in Figure \ref{fig:Figure4}a. While for the cross-polarized PC trace ($\sigma_{+} / \sigma_{-}$) in black a clear peak from the $X+$ transition is resolved, for the co-polarized PC trace ($\sigma_{+} / \sigma_{-}$) the amplitude of the $X_{+}$ transition in Figure \ref{fig:Figure4}a is completely suppressed due to Pauli blocking \cite{muller2012}. We estimate the initialization fidelity of the heavy hole spin to be $F_{\Uparrow} > 98.5 \%$ by extracting the integrated area $A$ of the PC absorption peaks of the positive trion $X_{+}$ for co- and cross-polarized excitation presented in Figure \ref{fig:Figure4}a and define the hole spin initialization fidelity as $F_{\Uparrow} = 1 - \frac{ A_{\sigma_{+} / \sigma_{+}}}{ A_{\sigma_{+} / \sigma_{+}} +  A_{\sigma_{+} / \sigma_{-}}}$ \cite{godden2012, godden2010}. Note, that the fidelity of the hole spin is mainly limited by the accuracy of fitting the area and the measurement noise and thus, we only give a lower bound for $F_{\Uparrow}$. Since the projection of the heavy hole spin on the z-axis scales with the $cosine$ of the exchange interaction precession angle, for spin initialization times $\tau_{e} = 0.6 ps$ is expected to exceed $F_{\Uparrow} > 98.5 \%$. 

To validate the ultrafast spin initialization time and demonstrate that the hole spin state $\Uparrow$ is stored with high fidelities, we present in Figure \ref{fig:Figure4}b the time dependence of the positively charged trion $X_{+}$ at $F=71.7 kV/cm$ for time delays between $\Delta t = 0 ns$ and $\Delta t = 0.5 ns$. The rise time $\tau_{e}$ of the cross-polarized amplitude encoded in black in Figure \ref{fig:Figure4}b, directly corresponds to the heavy hole spin initialization time and confirms the ultrafast initialization of the spin state $\Uparrow$. The constant suppression of the amplitude of the $X_{+}$ initialized and probed with $\sigma_{+} / \sigma_{+}$ polarized light presented in red in Figure \ref{fig:Figure4}b confirms that the heavy hole spin $\Uparrow$ is stored with a very high fidelity over the entire time range probed. In principle, by locking the excitation laser pulse to an electric field modulation, the Schottky diode structure can be switched to low electric field values after the heavy spin initialization allowing for arbitrarily long storage times. For the current device, we measure a voltage response time of $1.82ns$ for RC circuit model, that confirms that electric field switching within the hole storage time can easily be implemented (see supplementary material).  

\section{Conclusion}

In summary, we demonstrated the \textit{sub-picosecond} initialization of a single heavy hole spin without the need for a supporting external magnetic field with a fidelity of $F_{\Uparrow} > 98.5 \%$. Thereby, the heavy hole spin was initialized by tunnelling ionization of a coherently driven neutral exciton $X_{0}$. Using an $Al_{0.1}Ga_{0.9}As$ barrier adjacent to the QD we seperately control tunneling times of holes and electrons for (i) sub-picosecond hole spin initialization and (ii)simultaneous high-efficiency photocurrent read out of the QD charge and spin state. Embedding the QD layer in a Schottky diode structure allows for additional tuning of the charge carrier tunneling times by changing the applied electric field in growth direction. By mapping out field-dependent Rabi oscillations of the neutral $X_{0}$ and positively charged exciton $X_{+}$ transitions, we identified tunnelling induced dephasing of Rabi rotations as the major source for a strong intensity damping in this regime. Quantum optical simulations are in excellent agreement with the measurements and revealed that tunneling induced dephasing of Rabi oscillations can be used to extract the electron tunneling times from the excited state. This allowed to infer the change of Coulomb binding energy for having two holes in the QD. Most strikingly, strong tunneling induced dephasing of a neutral exciton transition results in the high-fidelity initialization of single hole states that is insensitive to errors in the excitation pulse intensity. In polarization resolved pump-probe measurements we demonstrated sub-picosecond hole spin initialization with near-unity fidelity and long hole storage times. The very fast and high fidelity initialization of single heavy hole spins makes the presented Schottky diode structure an ideal candidate for the realization of qubits and quantum protocols demanding high fidelity gate operations especially an ultra-high fidelity qubit initialization \cite{mcmahon2015}.

We gratefully acknowledge financial support from the DFG via SFB-631, Nanosystems Initiative Munich, the EU via S3 Nano and BaCaTeC. KM acknowledges financial support from the Alexander von Humboldt foundation and the ARO (grant W911NF-13-1-0309). KAF acknowledges financial support from the Lu Stanford Graduate Fellowship and the National Defense Science and Engineering Graduate Fellowship.

\bibliography{Papers}

\begin{thebibliography}{46}%
\makeatletter
\providecommand \@ifxundefined [1]{%
 \@ifx{#1\undefined}
}%
\providecommand \@ifnum [1]{%
 \ifnum #1\expandafter \@firstoftwo
 \else \expandafter \@secondoftwo
 \fi
}%
\providecommand \@ifx [1]{%
 \ifx #1\expandafter \@firstoftwo
 \else \expandafter \@secondoftwo
 \fi
}%
\providecommand \natexlab [1]{#1}%
\providecommand \enquote  [1]{``#1''}%
\providecommand \bibnamefont  [1]{#1}%
\providecommand \bibfnamefont [1]{#1}%
\providecommand \citenamefont [1]{#1}%
\providecommand \href@noop [0]{\@secondoftwo}%
\providecommand \href [0]{\begingroup \@sanitize@url \@href}%
\providecommand \@href[1]{\@@startlink{#1}\@@href}%
\providecommand \@@href[1]{\endgroup#1\@@endlink}%
\providecommand \@sanitize@url [0]{\catcode `\\12\catcode `\$12\catcode
  `\&12\catcode `\#12\catcode `\^12\catcode `\_12\catcode `\%12\relax}%
\providecommand \@@startlink[1]{}%
\providecommand \@@endlink[0]{}%
\providecommand \url  [0]{\begingroup\@sanitize@url \@url }%
\providecommand \@url [1]{\endgroup\@href {#1}{\urlprefix }}%
\providecommand \urlprefix  [0]{URL }%
\providecommand \Eprint [0]{\href }%
\providecommand \doibase [0]{http://dx.doi.org/}%
\providecommand \selectlanguage [0]{\@gobble}%
\providecommand \bibinfo  [0]{\@secondoftwo}%
\providecommand \bibfield  [0]{\@secondoftwo}%
\providecommand \translation [1]{[#1]}%
\providecommand \BibitemOpen [0]{}%
\providecommand \bibitemStop [0]{}%
\providecommand \bibitemNoStop [0]{.\EOS\space}%
\providecommand \EOS [0]{\spacefactor3000\relax}%
\providecommand \BibitemShut  [1]{\csname bibitem#1\endcsname}%
\let\auto@bib@innerbib\@empty
\bibitem [{\citenamefont {Loss}\ and\ \citenamefont
  {DiVincenzo}(1998)}]{loss1998}%
  \BibitemOpen
  \bibfield  {author} {\bibinfo {author} {\bibfnamefont {D.}~\bibnamefont
  {Loss}}\ and\ \bibinfo {author} {\bibfnamefont {D.~P.}\ \bibnamefont
  {DiVincenzo}},\ }\href@noop {} {\bibfield  {journal} {\bibinfo  {journal}
  {Physical Review A}\ }\textbf {\bibinfo {volume} {57}},\ \bibinfo {pages}
  {120} (\bibinfo {year} {1998})}\BibitemShut {NoStop}%
\bibitem [{\citenamefont {Kane}(1998)}]{kane1998}%
  \BibitemOpen
  \bibfield  {author} {\bibinfo {author} {\bibfnamefont {B.~E.}\ \bibnamefont
  {Kane}},\ }\href@noop {} {\bibfield  {journal} {\bibinfo  {journal} {nature}\
  }\textbf {\bibinfo {volume} {393}},\ \bibinfo {pages} {133} (\bibinfo {year}
  {1998})}\BibitemShut {NoStop}%
\bibitem [{\citenamefont {M{\"u}ller}\ \emph
  {et~al.}(2013{\natexlab{a}})\citenamefont {M{\"u}ller}, \citenamefont
  {Kaldewey}, \citenamefont {Ripszam}, \citenamefont {Wildmann}, \citenamefont
  {Bechtold}, \citenamefont {Bichler}, \citenamefont {Koblm{\"u}ller},
  \citenamefont {Abstreiter},\ and\ \citenamefont {Finley}}]{muller2013}%
  \BibitemOpen
  \bibfield  {author} {\bibinfo {author} {\bibfnamefont {K.}~\bibnamefont
  {M{\"u}ller}}, \bibinfo {author} {\bibfnamefont {T.}~\bibnamefont
  {Kaldewey}}, \bibinfo {author} {\bibfnamefont {R.}~\bibnamefont {Ripszam}},
  \bibinfo {author} {\bibfnamefont {J.}~\bibnamefont {Wildmann}}, \bibinfo
  {author} {\bibfnamefont {A.}~\bibnamefont {Bechtold}}, \bibinfo {author}
  {\bibfnamefont {M.}~\bibnamefont {Bichler}}, \bibinfo {author} {\bibfnamefont
  {G.}~\bibnamefont {Koblm{\"u}ller}}, \bibinfo {author} {\bibfnamefont
  {G.}~\bibnamefont {Abstreiter}}, \ and\ \bibinfo {author} {\bibfnamefont
  {J.}~\bibnamefont {Finley}},\ }\href@noop {} {\bibfield  {journal} {\bibinfo
  {journal} {Scientific reports}\ }\textbf {\bibinfo {volume} {3}} (\bibinfo
  {year} {2013}{\natexlab{a}})}\BibitemShut {NoStop}%
\bibitem [{\citenamefont {Press}\ \emph {et~al.}(2008)\citenamefont {Press},
  \citenamefont {Ladd}, \citenamefont {Zhang},\ and\ \citenamefont
  {Yamamoto}}]{press2008}%
  \BibitemOpen
  \bibfield  {author} {\bibinfo {author} {\bibfnamefont {D.}~\bibnamefont
  {Press}}, \bibinfo {author} {\bibfnamefont {T.~D.}\ \bibnamefont {Ladd}},
  \bibinfo {author} {\bibfnamefont {B.}~\bibnamefont {Zhang}}, \ and\ \bibinfo
  {author} {\bibfnamefont {Y.}~\bibnamefont {Yamamoto}},\ }\href {\doibase
  10.1038/nature07530} {\bibfield  {journal} {\bibinfo  {journal} {Nature}\
  }\textbf {\bibinfo {volume} {456}},\ \bibinfo {pages} {218} (\bibinfo {year}
  {2008})}\BibitemShut {NoStop}%
\bibitem [{\citenamefont {Press}\ \emph {et~al.}(2010)\citenamefont {Press},
  \citenamefont {De~Greve}, \citenamefont {McMahon}, \citenamefont {Ladd},
  \citenamefont {Friess}, \citenamefont {Schneider}, \citenamefont {Kamp},
  \citenamefont {H{\"o}fling}, \citenamefont {Forchel},\ and\ \citenamefont
  {Yamamoto}}]{press2010}%
  \BibitemOpen
  \bibfield  {author} {\bibinfo {author} {\bibfnamefont {D.}~\bibnamefont
  {Press}}, \bibinfo {author} {\bibfnamefont {K.}~\bibnamefont {De~Greve}},
  \bibinfo {author} {\bibfnamefont {P.~L.}\ \bibnamefont {McMahon}}, \bibinfo
  {author} {\bibfnamefont {T.~D.}\ \bibnamefont {Ladd}}, \bibinfo {author}
  {\bibfnamefont {B.}~\bibnamefont {Friess}}, \bibinfo {author} {\bibfnamefont
  {C.}~\bibnamefont {Schneider}}, \bibinfo {author} {\bibfnamefont
  {M.}~\bibnamefont {Kamp}}, \bibinfo {author} {\bibfnamefont {S.}~\bibnamefont
  {H{\"o}fling}}, \bibinfo {author} {\bibfnamefont {A.}~\bibnamefont
  {Forchel}}, \ and\ \bibinfo {author} {\bibfnamefont {Y.}~\bibnamefont
  {Yamamoto}},\ }\href@noop {} {\bibfield  {journal} {\bibinfo  {journal}
  {Nature Photonics}\ }\textbf {\bibinfo {volume} {4}},\ \bibinfo {pages} {367}
  (\bibinfo {year} {2010})}\BibitemShut {NoStop}%
\bibitem [{\citenamefont {De~Greve}\ \emph {et~al.}(2011)\citenamefont
  {De~Greve}, \citenamefont {McMahon}, \citenamefont {Press}, \citenamefont
  {Ladd}, \citenamefont {Bisping}, \citenamefont {Schneider}, \citenamefont
  {Kamp}, \citenamefont {Worschech}, \citenamefont {Hoefling}, \citenamefont
  {Forchel},\ and\ \citenamefont {Yamamoto}}]{DeGreve2011}%
  \BibitemOpen
  \bibfield  {author} {\bibinfo {author} {\bibfnamefont {K.}~\bibnamefont
  {De~Greve}}, \bibinfo {author} {\bibfnamefont {P.~L.}\ \bibnamefont
  {McMahon}}, \bibinfo {author} {\bibfnamefont {D.}~\bibnamefont {Press}},
  \bibinfo {author} {\bibfnamefont {T.~D.}\ \bibnamefont {Ladd}}, \bibinfo
  {author} {\bibfnamefont {D.}~\bibnamefont {Bisping}}, \bibinfo {author}
  {\bibfnamefont {C.}~\bibnamefont {Schneider}}, \bibinfo {author}
  {\bibfnamefont {M.}~\bibnamefont {Kamp}}, \bibinfo {author} {\bibfnamefont
  {L.}~\bibnamefont {Worschech}}, \bibinfo {author} {\bibfnamefont
  {S.}~\bibnamefont {Hoefling}}, \bibinfo {author} {\bibfnamefont
  {A.}~\bibnamefont {Forchel}}, \ and\ \bibinfo {author} {\bibfnamefont
  {Y.}~\bibnamefont {Yamamoto}},\ }\href {\doibase 10.1038/NPHYS2078}
  {\bibfield  {journal} {\bibinfo  {journal} {Nature Physics}\ }\textbf
  {\bibinfo {volume} {7}},\ \bibinfo {pages} {872} (\bibinfo {year}
  {2011})}\BibitemShut {NoStop}%
\bibitem [{\citenamefont {Vamivakas}\ \emph {et~al.}(2010)\citenamefont
  {Vamivakas}, \citenamefont {Lu}, \citenamefont {Matthiesen}, \citenamefont
  {Zhao}, \citenamefont {F{\"a}lt}, \citenamefont {Badolato},\ and\
  \citenamefont {Atat{\"u}re}}]{vamivakas2010}%
  \BibitemOpen
  \bibfield  {author} {\bibinfo {author} {\bibfnamefont {A.}~\bibnamefont
  {Vamivakas}}, \bibinfo {author} {\bibfnamefont {C.-Y.}\ \bibnamefont {Lu}},
  \bibinfo {author} {\bibfnamefont {C.}~\bibnamefont {Matthiesen}}, \bibinfo
  {author} {\bibfnamefont {Y.}~\bibnamefont {Zhao}}, \bibinfo {author}
  {\bibfnamefont {S.}~\bibnamefont {F{\"a}lt}}, \bibinfo {author}
  {\bibfnamefont {A.}~\bibnamefont {Badolato}}, \ and\ \bibinfo {author}
  {\bibfnamefont {M.}~\bibnamefont {Atat{\"u}re}},\ }\href@noop {} {\bibfield
  {journal} {\bibinfo  {journal} {Nature}\ }\textbf {\bibinfo {volume} {467}},\
  \bibinfo {pages} {297} (\bibinfo {year} {2010})}\BibitemShut {NoStop}%
\bibitem [{\citenamefont {Delteil}\ \emph {et~al.}(2014)\citenamefont
  {Delteil}, \citenamefont {Gao}, \citenamefont {Fallahi}, \citenamefont
  {Miguel-Sanchez},\ and\ \citenamefont {Imamo{\u{g}}lu}}]{delteil2014}%
  \BibitemOpen
  \bibfield  {author} {\bibinfo {author} {\bibfnamefont {A.}~\bibnamefont
  {Delteil}}, \bibinfo {author} {\bibfnamefont {W.-b.}\ \bibnamefont {Gao}},
  \bibinfo {author} {\bibfnamefont {P.}~\bibnamefont {Fallahi}}, \bibinfo
  {author} {\bibfnamefont {J.}~\bibnamefont {Miguel-Sanchez}}, \ and\ \bibinfo
  {author} {\bibfnamefont {A.}~\bibnamefont {Imamo{\u{g}}lu}},\ }\href@noop {}
  {\bibfield  {journal} {\bibinfo  {journal} {Physical review letters}\
  }\textbf {\bibinfo {volume} {112}},\ \bibinfo {pages} {116802} (\bibinfo
  {year} {2014})}\BibitemShut {NoStop}%
\bibitem [{\citenamefont {De~Greve}\ \emph {et~al.}(2012)\citenamefont
  {De~Greve}, \citenamefont {Yu}, \citenamefont {McMahon}, \citenamefont
  {Pelc}, \citenamefont {Natarajan}, \citenamefont {Kim}, \citenamefont {Abe},
  \citenamefont {Maier}, \citenamefont {Schneider}, \citenamefont {Kamp} \emph
  {et~al.}}]{yamamoto2012}%
  \BibitemOpen
  \bibfield  {author} {\bibinfo {author} {\bibfnamefont {K.}~\bibnamefont
  {De~Greve}}, \bibinfo {author} {\bibfnamefont {L.}~\bibnamefont {Yu}},
  \bibinfo {author} {\bibfnamefont {P.~L.}\ \bibnamefont {McMahon}}, \bibinfo
  {author} {\bibfnamefont {J.~S.}\ \bibnamefont {Pelc}}, \bibinfo {author}
  {\bibfnamefont {C.~M.}\ \bibnamefont {Natarajan}}, \bibinfo {author}
  {\bibfnamefont {N.~Y.}\ \bibnamefont {Kim}}, \bibinfo {author} {\bibfnamefont
  {E.}~\bibnamefont {Abe}}, \bibinfo {author} {\bibfnamefont {S.}~\bibnamefont
  {Maier}}, \bibinfo {author} {\bibfnamefont {C.}~\bibnamefont {Schneider}},
  \bibinfo {author} {\bibfnamefont {M.}~\bibnamefont {Kamp}},  \emph {et~al.},\
  }\href@noop {} {\bibfield  {journal} {\bibinfo  {journal} {Nature}\ }\textbf
  {\bibinfo {volume} {491}},\ \bibinfo {pages} {421} (\bibinfo {year}
  {2012})}\BibitemShut {NoStop}%
\bibitem [{\citenamefont {Gao}\ \emph {et~al.}(2012)\citenamefont {Gao},
  \citenamefont {Fallahi}, \citenamefont {Togan}, \citenamefont
  {Miguel-Sanchez},\ and\ \citenamefont {Imamoglu}}]{gao2012}%
  \BibitemOpen
  \bibfield  {author} {\bibinfo {author} {\bibfnamefont {W.}~\bibnamefont
  {Gao}}, \bibinfo {author} {\bibfnamefont {P.}~\bibnamefont {Fallahi}},
  \bibinfo {author} {\bibfnamefont {E.}~\bibnamefont {Togan}}, \bibinfo
  {author} {\bibfnamefont {J.}~\bibnamefont {Miguel-Sanchez}}, \ and\ \bibinfo
  {author} {\bibfnamefont {A.}~\bibnamefont {Imamoglu}},\ }\href@noop {}
  {\bibfield  {journal} {\bibinfo  {journal} {Nature}\ }\textbf {\bibinfo
  {volume} {491}},\ \bibinfo {pages} {426} (\bibinfo {year}
  {2012})}\BibitemShut {NoStop}%
\bibitem [{\citenamefont {Schaibley}\ \emph {et~al.}(2013)\citenamefont
  {Schaibley}, \citenamefont {Burgers}, \citenamefont {McCracken},
  \citenamefont {Duan}, \citenamefont {Berman}, \citenamefont {Steel},
  \citenamefont {Bracker}, \citenamefont {Gammon},\ and\ \citenamefont
  {Sham}}]{steel2013}%
  \BibitemOpen
  \bibfield  {author} {\bibinfo {author} {\bibfnamefont {J.}~\bibnamefont
  {Schaibley}}, \bibinfo {author} {\bibfnamefont {A.}~\bibnamefont {Burgers}},
  \bibinfo {author} {\bibfnamefont {G.}~\bibnamefont {McCracken}}, \bibinfo
  {author} {\bibfnamefont {L.-M.}\ \bibnamefont {Duan}}, \bibinfo {author}
  {\bibfnamefont {P.}~\bibnamefont {Berman}}, \bibinfo {author} {\bibfnamefont
  {D.}~\bibnamefont {Steel}}, \bibinfo {author} {\bibfnamefont
  {A.}~\bibnamefont {Bracker}}, \bibinfo {author} {\bibfnamefont
  {D.}~\bibnamefont {Gammon}}, \ and\ \bibinfo {author} {\bibfnamefont
  {L.}~\bibnamefont {Sham}},\ }\href@noop {} {\bibfield  {journal} {\bibinfo
  {journal} {Physical review letters}\ }\textbf {\bibinfo {volume} {110}},\
  \bibinfo {pages} {167401} (\bibinfo {year} {2013})}\BibitemShut {NoStop}%
\bibitem [{\citenamefont {McMahon}\ and\ \citenamefont
  {De~Greve}(2015)}]{mcmahon2015}%
  \BibitemOpen
  \bibfield  {author} {\bibinfo {author} {\bibfnamefont {P.~L.}\ \bibnamefont
  {McMahon}}\ and\ \bibinfo {author} {\bibfnamefont {K.}~\bibnamefont
  {De~Greve}},\ }\href@noop {} {\bibfield  {journal} {\bibinfo  {journal}
  {arXiv preprint arXiv:1501.03535}\ } (\bibinfo {year} {2015})}\BibitemShut
  {NoStop}%
\bibitem [{\citenamefont {M{\"u}ller}\ \emph
  {et~al.}(2012{\natexlab{a}})\citenamefont {M{\"u}ller}, \citenamefont
  {Bechtold}, \citenamefont {Ruppert}, \citenamefont {Hautmann}, \citenamefont
  {Wildmann}, \citenamefont {Kaldewey}, \citenamefont {Bichler}, \citenamefont
  {Krenner}, \citenamefont {Abstreiter}, \citenamefont {Betz} \emph
  {et~al.}}]{muller2012}%
  \BibitemOpen
  \bibfield  {author} {\bibinfo {author} {\bibfnamefont {K.}~\bibnamefont
  {M{\"u}ller}}, \bibinfo {author} {\bibfnamefont {A.}~\bibnamefont
  {Bechtold}}, \bibinfo {author} {\bibfnamefont {C.}~\bibnamefont {Ruppert}},
  \bibinfo {author} {\bibfnamefont {C.}~\bibnamefont {Hautmann}}, \bibinfo
  {author} {\bibfnamefont {J.}~\bibnamefont {Wildmann}}, \bibinfo {author}
  {\bibfnamefont {T.}~\bibnamefont {Kaldewey}}, \bibinfo {author}
  {\bibfnamefont {M.}~\bibnamefont {Bichler}}, \bibinfo {author} {\bibfnamefont
  {H.}~\bibnamefont {Krenner}}, \bibinfo {author} {\bibfnamefont
  {G.}~\bibnamefont {Abstreiter}}, \bibinfo {author} {\bibfnamefont
  {M.}~\bibnamefont {Betz}},  \emph {et~al.},\ }\href@noop {} {\bibfield
  {journal} {\bibinfo  {journal} {Physical Review B}\ }\textbf {\bibinfo
  {volume} {85}},\ \bibinfo {pages} {241306} (\bibinfo {year}
  {2012}{\natexlab{a}})}\BibitemShut {NoStop}%
\bibitem [{\citenamefont {Atat{\"u}re}\ \emph
  {et~al.}(2006{\natexlab{a}})\citenamefont {Atat{\"u}re}, \citenamefont
  {Dreiser}, \citenamefont {Badolato}, \citenamefont {H{\"o}gele},
  \citenamefont {Karrai},\ and\ \citenamefont {Imamoglu}}]{atature2006}%
  \BibitemOpen
  \bibfield  {author} {\bibinfo {author} {\bibfnamefont {M.}~\bibnamefont
  {Atat{\"u}re}}, \bibinfo {author} {\bibfnamefont {J.}~\bibnamefont
  {Dreiser}}, \bibinfo {author} {\bibfnamefont {A.}~\bibnamefont {Badolato}},
  \bibinfo {author} {\bibfnamefont {A.}~\bibnamefont {H{\"o}gele}}, \bibinfo
  {author} {\bibfnamefont {K.}~\bibnamefont {Karrai}}, \ and\ \bibinfo {author}
  {\bibfnamefont {A.}~\bibnamefont {Imamoglu}},\ }\href@noop {} {\bibfield
  {journal} {\bibinfo  {journal} {Science}\ }\textbf {\bibinfo {volume}
  {312}},\ \bibinfo {pages} {551} (\bibinfo {year}
  {2006}{\natexlab{a}})}\BibitemShut {NoStop}%
\bibitem [{\citenamefont {Gerardot}\ \emph {et~al.}(2008)\citenamefont
  {Gerardot}, \citenamefont {Brunner}, \citenamefont {Dalgarno}, \citenamefont
  {{\"O}hberg}, \citenamefont {Seidl}, \citenamefont {Kroner}, \citenamefont
  {Karrai}, \citenamefont {Stoltz}, \citenamefont {Petroff},\ and\
  \citenamefont {Warburton}}]{gerardot2008}%
  \BibitemOpen
  \bibfield  {author} {\bibinfo {author} {\bibfnamefont {B.~D.}\ \bibnamefont
  {Gerardot}}, \bibinfo {author} {\bibfnamefont {D.}~\bibnamefont {Brunner}},
  \bibinfo {author} {\bibfnamefont {P.~A.}\ \bibnamefont {Dalgarno}}, \bibinfo
  {author} {\bibfnamefont {P.}~\bibnamefont {{\"O}hberg}}, \bibinfo {author}
  {\bibfnamefont {S.}~\bibnamefont {Seidl}}, \bibinfo {author} {\bibfnamefont
  {M.}~\bibnamefont {Kroner}}, \bibinfo {author} {\bibfnamefont
  {K.}~\bibnamefont {Karrai}}, \bibinfo {author} {\bibfnamefont {N.~G.}\
  \bibnamefont {Stoltz}}, \bibinfo {author} {\bibfnamefont {P.~M.}\
  \bibnamefont {Petroff}}, \ and\ \bibinfo {author} {\bibfnamefont {R.~J.}\
  \bibnamefont {Warburton}},\ }\href@noop {} {\bibfield  {journal} {\bibinfo
  {journal} {Nature}\ }\textbf {\bibinfo {volume} {451}},\ \bibinfo {pages}
  {441} (\bibinfo {year} {2008})}\BibitemShut {NoStop}%
\bibitem [{\citenamefont {Kroutvar}\ \emph {et~al.}(2004)\citenamefont
  {Kroutvar}, \citenamefont {Ducommun}, \citenamefont {Heiss}, \citenamefont
  {Bichler}, \citenamefont {Schuh}, \citenamefont {Abstreiter},\ and\
  \citenamefont {Finley}}]{kroutvar2004}%
  \BibitemOpen
  \bibfield  {author} {\bibinfo {author} {\bibfnamefont {M.}~\bibnamefont
  {Kroutvar}}, \bibinfo {author} {\bibfnamefont {Y.}~\bibnamefont {Ducommun}},
  \bibinfo {author} {\bibfnamefont {D.}~\bibnamefont {Heiss}}, \bibinfo
  {author} {\bibfnamefont {M.}~\bibnamefont {Bichler}}, \bibinfo {author}
  {\bibfnamefont {D.}~\bibnamefont {Schuh}}, \bibinfo {author} {\bibfnamefont
  {G.}~\bibnamefont {Abstreiter}}, \ and\ \bibinfo {author} {\bibfnamefont
  {J.~J.}\ \bibnamefont {Finley}},\ }\href@noop {} {\bibfield  {journal}
  {\bibinfo  {journal} {Nature}\ }\textbf {\bibinfo {volume} {432}},\ \bibinfo
  {pages} {81} (\bibinfo {year} {2004})}\BibitemShut {NoStop}%
\bibitem [{\citenamefont {Heiss}\ \emph {et~al.}(2009)\citenamefont {Heiss},
  \citenamefont {Jovanov}, \citenamefont {Caesar}, \citenamefont {Bichler},
  \citenamefont {Abstreiter},\ and\ \citenamefont {Finley}}]{heiss2009}%
  \BibitemOpen
  \bibfield  {author} {\bibinfo {author} {\bibfnamefont {D.}~\bibnamefont
  {Heiss}}, \bibinfo {author} {\bibfnamefont {V.}~\bibnamefont {Jovanov}},
  \bibinfo {author} {\bibfnamefont {M.}~\bibnamefont {Caesar}}, \bibinfo
  {author} {\bibfnamefont {M.}~\bibnamefont {Bichler}}, \bibinfo {author}
  {\bibfnamefont {G.}~\bibnamefont {Abstreiter}}, \ and\ \bibinfo {author}
  {\bibfnamefont {J.}~\bibnamefont {Finley}},\ }\href@noop {} {\bibfield
  {journal} {\bibinfo  {journal} {Applied Physics Letters}\ }\textbf {\bibinfo
  {volume} {94}},\ \bibinfo {pages} {072108} (\bibinfo {year}
  {2009})}\BibitemShut {NoStop}%
\bibitem [{\citenamefont {Atat{\"u}re}\ \emph
  {et~al.}(2006{\natexlab{b}})\citenamefont {Atat{\"u}re}, \citenamefont
  {Dreiser}, \citenamefont {Badolato}, \citenamefont {H{\"o}gele},
  \citenamefont {Karrai},\ and\ \citenamefont {Imamoglu}}]{atature2006quantum}%
  \BibitemOpen
  \bibfield  {author} {\bibinfo {author} {\bibfnamefont {M.}~\bibnamefont
  {Atat{\"u}re}}, \bibinfo {author} {\bibfnamefont {J.}~\bibnamefont
  {Dreiser}}, \bibinfo {author} {\bibfnamefont {A.}~\bibnamefont {Badolato}},
  \bibinfo {author} {\bibfnamefont {A.}~\bibnamefont {H{\"o}gele}}, \bibinfo
  {author} {\bibfnamefont {K.}~\bibnamefont {Karrai}}, \ and\ \bibinfo {author}
  {\bibfnamefont {A.}~\bibnamefont {Imamoglu}},\ }\href@noop {} {\bibfield
  {journal} {\bibinfo  {journal} {Science}\ }\textbf {\bibinfo {volume}
  {312}},\ \bibinfo {pages} {551} (\bibinfo {year}
  {2006}{\natexlab{b}})}\BibitemShut {NoStop}%
\bibitem [{\citenamefont {Ramsay}\ \emph {et~al.}(2008)\citenamefont {Ramsay},
  \citenamefont {Boyle}, \citenamefont {Kolodka}, \citenamefont {Oliveira},
  \citenamefont {Skiba-Szymanska}, \citenamefont {Liu}, \citenamefont
  {Hopkinson}, \citenamefont {Fox},\ and\ \citenamefont
  {Skolnick}}]{ramsay2008fast}%
  \BibitemOpen
  \bibfield  {author} {\bibinfo {author} {\bibfnamefont {A.}~\bibnamefont
  {Ramsay}}, \bibinfo {author} {\bibfnamefont {S.}~\bibnamefont {Boyle}},
  \bibinfo {author} {\bibfnamefont {R.}~\bibnamefont {Kolodka}}, \bibinfo
  {author} {\bibfnamefont {J.~B. B.~d.}\ \bibnamefont {Oliveira}}, \bibinfo
  {author} {\bibfnamefont {J.}~\bibnamefont {Skiba-Szymanska}}, \bibinfo
  {author} {\bibfnamefont {H.}~\bibnamefont {Liu}}, \bibinfo {author}
  {\bibfnamefont {M.}~\bibnamefont {Hopkinson}}, \bibinfo {author}
  {\bibfnamefont {A.}~\bibnamefont {Fox}}, \ and\ \bibinfo {author}
  {\bibfnamefont {M.}~\bibnamefont {Skolnick}},\ }\href@noop {} {\bibfield
  {journal} {\bibinfo  {journal} {Physical review letters}\ }\textbf {\bibinfo
  {volume} {100}},\ \bibinfo {pages} {197401} (\bibinfo {year}
  {2008})}\BibitemShut {NoStop}%
\bibitem [{\citenamefont {Godden}\ \emph
  {et~al.}(2012{\natexlab{a}})\citenamefont {Godden}, \citenamefont {Quilter},
  \citenamefont {Ramsay}, \citenamefont {Wu}, \citenamefont {Brereton},
  \citenamefont {Boyle}, \citenamefont {Luxmoore}, \citenamefont
  {Puebla-Nunez}, \citenamefont {Fox},\ and\ \citenamefont
  {Skolnick}}]{godden2012}%
  \BibitemOpen
  \bibfield  {author} {\bibinfo {author} {\bibfnamefont {T.}~\bibnamefont
  {Godden}}, \bibinfo {author} {\bibfnamefont {J.}~\bibnamefont {Quilter}},
  \bibinfo {author} {\bibfnamefont {A.}~\bibnamefont {Ramsay}}, \bibinfo
  {author} {\bibfnamefont {Y.}~\bibnamefont {Wu}}, \bibinfo {author}
  {\bibfnamefont {P.}~\bibnamefont {Brereton}}, \bibinfo {author}
  {\bibfnamefont {S.}~\bibnamefont {Boyle}}, \bibinfo {author} {\bibfnamefont
  {I.}~\bibnamefont {Luxmoore}}, \bibinfo {author} {\bibfnamefont
  {J.}~\bibnamefont {Puebla-Nunez}}, \bibinfo {author} {\bibfnamefont
  {A.}~\bibnamefont {Fox}}, \ and\ \bibinfo {author} {\bibfnamefont
  {M.}~\bibnamefont {Skolnick}},\ }\href@noop {} {\bibfield  {journal}
  {\bibinfo  {journal} {Physical review letters}\ }\textbf {\bibinfo {volume}
  {108}},\ \bibinfo {pages} {017402} (\bibinfo {year}
  {2012}{\natexlab{a}})}\BibitemShut {NoStop}%
\bibitem [{\citenamefont {Godden}\ \emph
  {et~al.}(2012{\natexlab{b}})\citenamefont {Godden}, \citenamefont {Quilter},
  \citenamefont {Ramsay}, \citenamefont {Wu}, \citenamefont {Brereton},
  \citenamefont {Luxmoore}, \citenamefont {Puebla}, \citenamefont {Fox},\ and\
  \citenamefont {Skolnick}}]{godden2012fast}%
  \BibitemOpen
  \bibfield  {author} {\bibinfo {author} {\bibfnamefont {T.}~\bibnamefont
  {Godden}}, \bibinfo {author} {\bibfnamefont {J.}~\bibnamefont {Quilter}},
  \bibinfo {author} {\bibfnamefont {A.}~\bibnamefont {Ramsay}}, \bibinfo
  {author} {\bibfnamefont {Y.}~\bibnamefont {Wu}}, \bibinfo {author}
  {\bibfnamefont {P.}~\bibnamefont {Brereton}}, \bibinfo {author}
  {\bibfnamefont {I.}~\bibnamefont {Luxmoore}}, \bibinfo {author}
  {\bibfnamefont {J.}~\bibnamefont {Puebla}}, \bibinfo {author} {\bibfnamefont
  {A.}~\bibnamefont {Fox}}, \ and\ \bibinfo {author} {\bibfnamefont
  {M.}~\bibnamefont {Skolnick}},\ }\href@noop {} {\bibfield  {journal}
  {\bibinfo  {journal} {Physical Review B}\ }\textbf {\bibinfo {volume} {85}},\
  \bibinfo {pages} {155310} (\bibinfo {year} {2012}{\natexlab{b}})}\BibitemShut
  {NoStop}%
\bibitem [{\citenamefont {Mar}\ \emph {et~al.}(2014)\citenamefont {Mar},
  \citenamefont {Baumberg}, \citenamefont {Xu}, \citenamefont {Irvine},\ and\
  \citenamefont {Williams}}]{mar2014ultrafast}%
  \BibitemOpen
  \bibfield  {author} {\bibinfo {author} {\bibfnamefont {J.~D.}\ \bibnamefont
  {Mar}}, \bibinfo {author} {\bibfnamefont {J.~J.}\ \bibnamefont {Baumberg}},
  \bibinfo {author} {\bibfnamefont {X.}~\bibnamefont {Xu}}, \bibinfo {author}
  {\bibfnamefont {A.~C.}\ \bibnamefont {Irvine}}, \ and\ \bibinfo {author}
  {\bibfnamefont {D.~A.}\ \bibnamefont {Williams}},\ }\href@noop {} {\bibfield
  {journal} {\bibinfo  {journal} {Physical Review B}\ }\textbf {\bibinfo
  {volume} {90}},\ \bibinfo {pages} {241303} (\bibinfo {year}
  {2014})}\BibitemShut {NoStop}%
\bibitem [{\citenamefont {Oulton}\ \emph
  {et~al.}(2002{\natexlab{a}})\citenamefont {Oulton}, \citenamefont {Finley},
  \citenamefont {Ashmore}, \citenamefont {Gregory}, \citenamefont {Mowbray},
  \citenamefont {Skolnick}, \citenamefont {Steer}, \citenamefont {Liew},
  \citenamefont {Migliorato},\ and\ \citenamefont
  {Cullis}}]{oulton2002manipulation}%
  \BibitemOpen
  \bibfield  {author} {\bibinfo {author} {\bibfnamefont {R.}~\bibnamefont
  {Oulton}}, \bibinfo {author} {\bibfnamefont {J.}~\bibnamefont {Finley}},
  \bibinfo {author} {\bibfnamefont {A.}~\bibnamefont {Ashmore}}, \bibinfo
  {author} {\bibfnamefont {I.}~\bibnamefont {Gregory}}, \bibinfo {author}
  {\bibfnamefont {D.}~\bibnamefont {Mowbray}}, \bibinfo {author} {\bibfnamefont
  {M.}~\bibnamefont {Skolnick}}, \bibinfo {author} {\bibfnamefont
  {M.}~\bibnamefont {Steer}}, \bibinfo {author} {\bibfnamefont {S.-L.}\
  \bibnamefont {Liew}}, \bibinfo {author} {\bibfnamefont {M.}~\bibnamefont
  {Migliorato}}, \ and\ \bibinfo {author} {\bibfnamefont {A.}~\bibnamefont
  {Cullis}},\ }\href@noop {} {\bibfield  {journal} {\bibinfo  {journal}
  {Physical Review B}\ }\textbf {\bibinfo {volume} {66}},\ \bibinfo {pages}
  {045313} (\bibinfo {year} {2002}{\natexlab{a}})}\BibitemShut {NoStop}%
\bibitem [{\citenamefont {Godden}\ \emph {et~al.}(2010)\citenamefont {Godden},
  \citenamefont {Boyle}, \citenamefont {Ramsay}, \citenamefont {Fox},\ and\
  \citenamefont {Skolnick}}]{godden2010}%
  \BibitemOpen
  \bibfield  {author} {\bibinfo {author} {\bibfnamefont {T.~M.}\ \bibnamefont
  {Godden}}, \bibinfo {author} {\bibfnamefont {S.~J.}\ \bibnamefont {Boyle}},
  \bibinfo {author} {\bibfnamefont {A.~J.}\ \bibnamefont {Ramsay}}, \bibinfo
  {author} {\bibfnamefont {A.}~\bibnamefont {Fox}}, \ and\ \bibinfo {author}
  {\bibfnamefont {M.}~\bibnamefont {Skolnick}},\ }\href@noop {} {\bibfield
  {journal} {\bibinfo  {journal} {Applied Physics Letters}\ }\textbf {\bibinfo
  {volume} {97}},\ \bibinfo {pages} {061113} (\bibinfo {year}
  {2010})}\BibitemShut {NoStop}%
\bibitem [{\citenamefont {Warburton}\ \emph {et~al.}(2000)\citenamefont
  {Warburton}, \citenamefont {Sch{\"a}flein}, \citenamefont {Haft},
  \citenamefont {Bickel}, \citenamefont {Lorke}, \citenamefont {Karrai},
  \citenamefont {Garcia}, \citenamefont {Schoenfeld},\ and\ \citenamefont
  {Petroff}}]{warburton2000}%
  \BibitemOpen
  \bibfield  {author} {\bibinfo {author} {\bibfnamefont {R.~J.}\ \bibnamefont
  {Warburton}}, \bibinfo {author} {\bibfnamefont {C.}~\bibnamefont
  {Sch{\"a}flein}}, \bibinfo {author} {\bibfnamefont {D.}~\bibnamefont {Haft}},
  \bibinfo {author} {\bibfnamefont {F.}~\bibnamefont {Bickel}}, \bibinfo
  {author} {\bibfnamefont {A.}~\bibnamefont {Lorke}}, \bibinfo {author}
  {\bibfnamefont {K.}~\bibnamefont {Karrai}}, \bibinfo {author} {\bibfnamefont
  {J.~M.}\ \bibnamefont {Garcia}}, \bibinfo {author} {\bibfnamefont
  {W.}~\bibnamefont {Schoenfeld}}, \ and\ \bibinfo {author} {\bibfnamefont
  {P.~M.}\ \bibnamefont {Petroff}},\ }\href@noop {} {\bibfield  {journal}
  {\bibinfo  {journal} {Nature}\ }\textbf {\bibinfo {volume} {405}},\ \bibinfo
  {pages} {926} (\bibinfo {year} {2000})}\BibitemShut {NoStop}%
\bibitem [{\citenamefont {Krenner}\ \emph {et~al.}(2005)\citenamefont
  {Krenner}, \citenamefont {Stufler}, \citenamefont {Sabathil}, \citenamefont
  {Clark}, \citenamefont {Ester}, \citenamefont {Bichler}, \citenamefont
  {Abstreiter}, \citenamefont {Finley},\ and\ \citenamefont
  {Zrenner}}]{Krenner2005}%
  \BibitemOpen
  \bibfield  {author} {\bibinfo {author} {\bibfnamefont {H.~J.}\ \bibnamefont
  {Krenner}}, \bibinfo {author} {\bibfnamefont {S.}~\bibnamefont {Stufler}},
  \bibinfo {author} {\bibfnamefont {M.}~\bibnamefont {Sabathil}}, \bibinfo
  {author} {\bibfnamefont {E.~C.}\ \bibnamefont {Clark}}, \bibinfo {author}
  {\bibfnamefont {P.}~\bibnamefont {Ester}}, \bibinfo {author} {\bibfnamefont
  {M.}~\bibnamefont {Bichler}}, \bibinfo {author} {\bibfnamefont
  {G.}~\bibnamefont {Abstreiter}}, \bibinfo {author} {\bibfnamefont {J.~J.}\
  \bibnamefont {Finley}}, \ and\ \bibinfo {author} {\bibfnamefont
  {A.}~\bibnamefont {Zrenner}},\ }\href@noop {} {\bibfield  {journal} {\bibinfo
   {journal} {New Journal of Physics}\ }\textbf {\bibinfo {volume} {7}},\
  \bibinfo {pages} {184} (\bibinfo {year} {2005})}\BibitemShut {NoStop}%
\bibitem [{\citenamefont {Zrenner}\ \emph {et~al.}(2002)\citenamefont
  {Zrenner}, \citenamefont {Beham}, \citenamefont {Stufler}, \citenamefont
  {Findeis}, \citenamefont {Bichler},\ and\ \citenamefont
  {Abstreiter}}]{zrenner2002}%
  \BibitemOpen
  \bibfield  {author} {\bibinfo {author} {\bibfnamefont {A.}~\bibnamefont
  {Zrenner}}, \bibinfo {author} {\bibfnamefont {E.}~\bibnamefont {Beham}},
  \bibinfo {author} {\bibfnamefont {S.}~\bibnamefont {Stufler}}, \bibinfo
  {author} {\bibfnamefont {F.}~\bibnamefont {Findeis}}, \bibinfo {author}
  {\bibfnamefont {M.}~\bibnamefont {Bichler}}, \ and\ \bibinfo {author}
  {\bibfnamefont {G.}~\bibnamefont {Abstreiter}},\ }\href@noop {} {\bibfield
  {journal} {\bibinfo  {journal} {Nature}\ }\textbf {\bibinfo {volume} {418}},\
  \bibinfo {pages} {612} (\bibinfo {year} {2002})}\BibitemShut {NoStop}%
\bibitem [{\citenamefont {M{\"u}ller}\ \emph
  {et~al.}(2013{\natexlab{b}})\citenamefont {M{\"u}ller}, \citenamefont
  {Bechtold}, \citenamefont {Ruppert}, \citenamefont {Kaldewey}, \citenamefont
  {Zecherle}, \citenamefont {Wildmann}, \citenamefont {Bichler}, \citenamefont
  {Krenner}, \citenamefont {Villas-B{\^o}as}, \citenamefont {Abstreiter} \emph
  {et~al.}}]{muller2013probing}%
  \BibitemOpen
  \bibfield  {author} {\bibinfo {author} {\bibfnamefont {K.}~\bibnamefont
  {M{\"u}ller}}, \bibinfo {author} {\bibfnamefont {A.}~\bibnamefont
  {Bechtold}}, \bibinfo {author} {\bibfnamefont {C.}~\bibnamefont {Ruppert}},
  \bibinfo {author} {\bibfnamefont {T.}~\bibnamefont {Kaldewey}}, \bibinfo
  {author} {\bibfnamefont {M.}~\bibnamefont {Zecherle}}, \bibinfo {author}
  {\bibfnamefont {J.~S.}\ \bibnamefont {Wildmann}}, \bibinfo {author}
  {\bibfnamefont {M.}~\bibnamefont {Bichler}}, \bibinfo {author} {\bibfnamefont
  {H.~J.}\ \bibnamefont {Krenner}}, \bibinfo {author} {\bibfnamefont {J.~M.}\
  \bibnamefont {Villas-B{\^o}as}}, \bibinfo {author} {\bibfnamefont
  {G.}~\bibnamefont {Abstreiter}},  \emph {et~al.},\ }\href@noop {} {\bibfield
  {journal} {\bibinfo  {journal} {Annalen der Physik}\ }\textbf {\bibinfo
  {volume} {525}},\ \bibinfo {pages} {49} (\bibinfo {year}
  {2013}{\natexlab{b}})}\BibitemShut {NoStop}%
\bibitem [{\citenamefont {Godden}\ \emph
  {et~al.}(2012{\natexlab{c}})\citenamefont {Godden}, \citenamefont {Quilter},
  \citenamefont {Ramsay}, \citenamefont {Wu}, \citenamefont {Brereton},
  \citenamefont {Boyle}, \citenamefont {Luxmoore}, \citenamefont
  {Puebla-Nunez}, \citenamefont {Fox},\ and\ \citenamefont
  {Skolnick}}]{godden2012coherent}%
  \BibitemOpen
  \bibfield  {author} {\bibinfo {author} {\bibfnamefont {T.}~\bibnamefont
  {Godden}}, \bibinfo {author} {\bibfnamefont {J.}~\bibnamefont {Quilter}},
  \bibinfo {author} {\bibfnamefont {A.}~\bibnamefont {Ramsay}}, \bibinfo
  {author} {\bibfnamefont {Y.}~\bibnamefont {Wu}}, \bibinfo {author}
  {\bibfnamefont {P.}~\bibnamefont {Brereton}}, \bibinfo {author}
  {\bibfnamefont {S.}~\bibnamefont {Boyle}}, \bibinfo {author} {\bibfnamefont
  {I.}~\bibnamefont {Luxmoore}}, \bibinfo {author} {\bibfnamefont
  {J.}~\bibnamefont {Puebla-Nunez}}, \bibinfo {author} {\bibfnamefont
  {A.}~\bibnamefont {Fox}}, \ and\ \bibinfo {author} {\bibfnamefont
  {M.}~\bibnamefont {Skolnick}},\ }\href@noop {} {\bibfield  {journal}
  {\bibinfo  {journal} {Physical review letters}\ }\textbf {\bibinfo {volume}
  {108}},\ \bibinfo {pages} {017402} (\bibinfo {year}
  {2012}{\natexlab{c}})}\BibitemShut {NoStop}%
\bibitem [{\citenamefont {M{\"u}ller}\ \emph
  {et~al.}(2012{\natexlab{b}})\citenamefont {M{\"u}ller}, \citenamefont
  {Bechtold}, \citenamefont {Ruppert}, \citenamefont {Zecherle}, \citenamefont
  {Reithmaier}, \citenamefont {Bichler}, \citenamefont {Krenner}, \citenamefont
  {Abstreiter}, \citenamefont {Holleitner}, \citenamefont {Villas-Boas} \emph
  {et~al.}}]{muller2012phonon}%
  \BibitemOpen
  \bibfield  {author} {\bibinfo {author} {\bibfnamefont {K.}~\bibnamefont
  {M{\"u}ller}}, \bibinfo {author} {\bibfnamefont {A.}~\bibnamefont
  {Bechtold}}, \bibinfo {author} {\bibfnamefont {C.}~\bibnamefont {Ruppert}},
  \bibinfo {author} {\bibfnamefont {M.}~\bibnamefont {Zecherle}}, \bibinfo
  {author} {\bibfnamefont {G.}~\bibnamefont {Reithmaier}}, \bibinfo {author}
  {\bibfnamefont {M.}~\bibnamefont {Bichler}}, \bibinfo {author} {\bibfnamefont
  {H.}~\bibnamefont {Krenner}}, \bibinfo {author} {\bibfnamefont
  {G.}~\bibnamefont {Abstreiter}}, \bibinfo {author} {\bibfnamefont
  {A.}~\bibnamefont {Holleitner}}, \bibinfo {author} {\bibfnamefont
  {J.}~\bibnamefont {Villas-Boas}},  \emph {et~al.},\ }\href@noop {} {\bibfield
   {journal} {\bibinfo  {journal} {Physical review letters}\ }\textbf {\bibinfo
  {volume} {108}},\ \bibinfo {pages} {197402} (\bibinfo {year}
  {2012}{\natexlab{b}})}\BibitemShut {NoStop}%
\bibitem [{\citenamefont {Goldberg}\ and\ \citenamefont
  {Schmidt}(1999)}]{goldberg1999handbook}%
  \BibitemOpen
  \bibfield  {author} {\bibinfo {author} {\bibfnamefont {Y.~A.}\ \bibnamefont
  {Goldberg}}\ and\ \bibinfo {author} {\bibfnamefont {N.}~\bibnamefont
  {Schmidt}},\ }\href@noop {} {\bibfield  {journal} {\bibinfo  {journal} {vol}\
  }\textbf {\bibinfo {volume} {1}},\ \bibinfo {pages} {191} (\bibinfo {year}
  {1999})}\BibitemShut {NoStop}%
\bibitem [{\citenamefont {Fry}\ \emph {et~al.}(2000)\citenamefont {Fry},
  \citenamefont {Finley}, \citenamefont {Wilson}, \citenamefont {Lemaitre},
  \citenamefont {Mowbray}, \citenamefont {Skolnick}, \citenamefont {Hopkinson},
  \citenamefont {Hill},\ and\ \citenamefont {Clark}}]{fry2000}%
  \BibitemOpen
  \bibfield  {author} {\bibinfo {author} {\bibfnamefont {P.}~\bibnamefont
  {Fry}}, \bibinfo {author} {\bibfnamefont {J.}~\bibnamefont {Finley}},
  \bibinfo {author} {\bibfnamefont {L.}~\bibnamefont {Wilson}}, \bibinfo
  {author} {\bibfnamefont {A.}~\bibnamefont {Lemaitre}}, \bibinfo {author}
  {\bibfnamefont {D.}~\bibnamefont {Mowbray}}, \bibinfo {author} {\bibfnamefont
  {M.}~\bibnamefont {Skolnick}}, \bibinfo {author} {\bibfnamefont
  {M.}~\bibnamefont {Hopkinson}}, \bibinfo {author} {\bibfnamefont
  {G.}~\bibnamefont {Hill}}, \ and\ \bibinfo {author} {\bibfnamefont
  {J.}~\bibnamefont {Clark}},\ }\href@noop {} {\bibfield  {journal} {\bibinfo
  {journal} {Applied Physics Letters}\ }\textbf {\bibinfo {volume} {77}},\
  \bibinfo {pages} {4344} (\bibinfo {year} {2000})}\BibitemShut {NoStop}%
\bibitem [{\citenamefont {Villas-B{\^o}as}\ \emph {et~al.}(2005)\citenamefont
  {Villas-B{\^o}as}, \citenamefont {Ulloa},\ and\ \citenamefont
  {Govorov}}]{villas2005}%
  \BibitemOpen
  \bibfield  {author} {\bibinfo {author} {\bibfnamefont {J.}~\bibnamefont
  {Villas-B{\^o}as}}, \bibinfo {author} {\bibfnamefont {S.~E.}\ \bibnamefont
  {Ulloa}}, \ and\ \bibinfo {author} {\bibfnamefont {A.}~\bibnamefont
  {Govorov}},\ }\href@noop {} {\bibfield  {journal} {\bibinfo  {journal}
  {Physical review letters}\ }\textbf {\bibinfo {volume} {94}},\ \bibinfo
  {pages} {057404} (\bibinfo {year} {2005})}\BibitemShut {NoStop}%
\bibitem [{\citenamefont {Ramsay}\ \emph
  {et~al.}(2010{\natexlab{a}})\citenamefont {Ramsay}, \citenamefont {Gopal},
  \citenamefont {Gauger}, \citenamefont {Nazir}, \citenamefont {Lovett},
  \citenamefont {Fox},\ and\ \citenamefont {Skolnick}}]{ramsay2010damping}%
  \BibitemOpen
  \bibfield  {author} {\bibinfo {author} {\bibfnamefont {A.}~\bibnamefont
  {Ramsay}}, \bibinfo {author} {\bibfnamefont {A.~V.}\ \bibnamefont {Gopal}},
  \bibinfo {author} {\bibfnamefont {E.}~\bibnamefont {Gauger}}, \bibinfo
  {author} {\bibfnamefont {A.}~\bibnamefont {Nazir}}, \bibinfo {author}
  {\bibfnamefont {B.}~\bibnamefont {Lovett}}, \bibinfo {author} {\bibfnamefont
  {A.}~\bibnamefont {Fox}}, \ and\ \bibinfo {author} {\bibfnamefont
  {M.}~\bibnamefont {Skolnick}},\ }\href@noop {} {\bibfield  {journal}
  {\bibinfo  {journal} {Physical review letters}\ }\textbf {\bibinfo {volume}
  {104}},\ \bibinfo {pages} {017402} (\bibinfo {year}
  {2010}{\natexlab{a}})}\BibitemShut {NoStop}%
\bibitem [{\citenamefont {Ardelt}\ \emph {et~al.}(2014)\citenamefont {Ardelt},
  \citenamefont {Hanschke}, \citenamefont {Fischer}, \citenamefont {M\"uller},
  \citenamefont {Kleinkauf}, \citenamefont {Koller}, \citenamefont {Bechtold},
  \citenamefont {Simmet}, \citenamefont {Wierzbowski}, \citenamefont {Riedl},
  \citenamefont {Abstreiter},\ and\ \citenamefont {Finley}}]{ardelt}%
  \BibitemOpen
  \bibfield  {author} {\bibinfo {author} {\bibfnamefont {P.-L.}\ \bibnamefont
  {Ardelt}}, \bibinfo {author} {\bibfnamefont {L.}~\bibnamefont {Hanschke}},
  \bibinfo {author} {\bibfnamefont {K.~A.}\ \bibnamefont {Fischer}}, \bibinfo
  {author} {\bibfnamefont {K.}~\bibnamefont {M\"uller}}, \bibinfo {author}
  {\bibfnamefont {A.}~\bibnamefont {Kleinkauf}}, \bibinfo {author}
  {\bibfnamefont {M.}~\bibnamefont {Koller}}, \bibinfo {author} {\bibfnamefont
  {A.}~\bibnamefont {Bechtold}}, \bibinfo {author} {\bibfnamefont
  {T.}~\bibnamefont {Simmet}}, \bibinfo {author} {\bibfnamefont
  {J.}~\bibnamefont {Wierzbowski}}, \bibinfo {author} {\bibfnamefont
  {H.}~\bibnamefont {Riedl}}, \bibinfo {author} {\bibfnamefont
  {G.}~\bibnamefont {Abstreiter}}, \ and\ \bibinfo {author} {\bibfnamefont
  {J.~J.}\ \bibnamefont {Finley}},\ }\href {\doibase
  10.1103/PhysRevB.90.241404} {\bibfield  {journal} {\bibinfo  {journal} {Phys.
  Rev. B}\ }\textbf {\bibinfo {volume} {90}},\ \bibinfo {pages} {241404}
  (\bibinfo {year} {2014})}\BibitemShut {NoStop}%
\bibitem [{\citenamefont {Ramsay}\ \emph
  {et~al.}(2010{\natexlab{b}})\citenamefont {Ramsay}, \citenamefont {Godden},
  \citenamefont {Boyle}, \citenamefont {Gauger}, \citenamefont {Nazir},
  \citenamefont {Lovett}, \citenamefont {Fox},\ and\ \citenamefont
  {Skolnick}}]{ramsay2010phonon}%
  \BibitemOpen
  \bibfield  {author} {\bibinfo {author} {\bibfnamefont {A.}~\bibnamefont
  {Ramsay}}, \bibinfo {author} {\bibfnamefont {T.}~\bibnamefont {Godden}},
  \bibinfo {author} {\bibfnamefont {S.}~\bibnamefont {Boyle}}, \bibinfo
  {author} {\bibfnamefont {E.~M.}\ \bibnamefont {Gauger}}, \bibinfo {author}
  {\bibfnamefont {A.}~\bibnamefont {Nazir}}, \bibinfo {author} {\bibfnamefont
  {B.~W.}\ \bibnamefont {Lovett}}, \bibinfo {author} {\bibfnamefont
  {A.}~\bibnamefont {Fox}}, \ and\ \bibinfo {author} {\bibfnamefont
  {M.}~\bibnamefont {Skolnick}},\ }\href@noop {} {\bibfield  {journal}
  {\bibinfo  {journal} {Physical review letters}\ }\textbf {\bibinfo {volume}
  {105}},\ \bibinfo {pages} {177402} (\bibinfo {year}
  {2010}{\natexlab{b}})}\BibitemShut {NoStop}%
\bibitem [{\citenamefont {Johansson}\ \emph {et~al.}(2012)\citenamefont
  {Johansson}, \citenamefont {Nation},\ and\ \citenamefont {Nori}}]{qutip}%
  \BibitemOpen
  \bibfield  {author} {\bibinfo {author} {\bibfnamefont {J.}~\bibnamefont
  {Johansson}}, \bibinfo {author} {\bibfnamefont {P.}~\bibnamefont {Nation}}, \
  and\ \bibinfo {author} {\bibfnamefont {F.}~\bibnamefont {Nori}},\ }\href@noop
  {} {\bibfield  {journal} {\bibinfo  {journal} {Computer Physics
  Communications}\ }\textbf {\bibinfo {volume} {183}},\ \bibinfo {pages} {1760}
  (\bibinfo {year} {2012})}\BibitemShut {NoStop}%
\bibitem [{\citenamefont {Finley}\ \emph {et~al.}(2004)\citenamefont {Finley},
  \citenamefont {Sabathil}, \citenamefont {Vogl}, \citenamefont {Abstreiter},
  \citenamefont {Oulton}, \citenamefont {Tartakovskii}, \citenamefont
  {Mowbray}, \citenamefont {Skolnick}, \citenamefont {Liew}, \citenamefont
  {Cullis} \emph {et~al.}}]{finley2004quantum}%
  \BibitemOpen
  \bibfield  {author} {\bibinfo {author} {\bibfnamefont {J.}~\bibnamefont
  {Finley}}, \bibinfo {author} {\bibfnamefont {M.}~\bibnamefont {Sabathil}},
  \bibinfo {author} {\bibfnamefont {P.}~\bibnamefont {Vogl}}, \bibinfo {author}
  {\bibfnamefont {G.}~\bibnamefont {Abstreiter}}, \bibinfo {author}
  {\bibfnamefont {R.}~\bibnamefont {Oulton}}, \bibinfo {author} {\bibfnamefont
  {A.}~\bibnamefont {Tartakovskii}}, \bibinfo {author} {\bibfnamefont
  {D.}~\bibnamefont {Mowbray}}, \bibinfo {author} {\bibfnamefont
  {M.}~\bibnamefont {Skolnick}}, \bibinfo {author} {\bibfnamefont
  {S.}~\bibnamefont {Liew}}, \bibinfo {author} {\bibfnamefont {A.}~\bibnamefont
  {Cullis}},  \emph {et~al.},\ }\href@noop {} {\bibfield  {journal} {\bibinfo
  {journal} {Physical Review B}\ }\textbf {\bibinfo {volume} {70}},\ \bibinfo
  {pages} {201308} (\bibinfo {year} {2004})}\BibitemShut {NoStop}%
\bibitem [{\citenamefont {Muller}\ \emph {et~al.}(2007)\citenamefont {Muller},
  \citenamefont {Flagg}, \citenamefont {Bianucci}, \citenamefont {Wang},
  \citenamefont {Deppe}, \citenamefont {Ma}, \citenamefont {Zhang},
  \citenamefont {Salamo}, \citenamefont {Xiao},\ and\ \citenamefont
  {Shih}}]{muller2007}%
  \BibitemOpen
  \bibfield  {author} {\bibinfo {author} {\bibfnamefont {A.}~\bibnamefont
  {Muller}}, \bibinfo {author} {\bibfnamefont {E.~B.}\ \bibnamefont {Flagg}},
  \bibinfo {author} {\bibfnamefont {P.}~\bibnamefont {Bianucci}}, \bibinfo
  {author} {\bibfnamefont {X.}~\bibnamefont {Wang}}, \bibinfo {author}
  {\bibfnamefont {D.~G.}\ \bibnamefont {Deppe}}, \bibinfo {author}
  {\bibfnamefont {W.}~\bibnamefont {Ma}}, \bibinfo {author} {\bibfnamefont
  {J.}~\bibnamefont {Zhang}}, \bibinfo {author} {\bibfnamefont
  {G.}~\bibnamefont {Salamo}}, \bibinfo {author} {\bibfnamefont
  {M.}~\bibnamefont {Xiao}}, \ and\ \bibinfo {author} {\bibfnamefont {C.-K.}\
  \bibnamefont {Shih}},\ }\href@noop {} {\bibfield  {journal} {\bibinfo
  {journal} {Physical Review Letters}\ }\textbf {\bibinfo {volume} {99}},\
  \bibinfo {pages} {187402} (\bibinfo {year} {2007})}\BibitemShut {NoStop}%
\bibitem [{\citenamefont {Matthiesen}\ \emph {et~al.}(2012)\citenamefont
  {Matthiesen}, \citenamefont {Vamivakas},\ and\ \citenamefont
  {Atat{\"u}re}}]{matthiesen2012}%
  \BibitemOpen
  \bibfield  {author} {\bibinfo {author} {\bibfnamefont {C.}~\bibnamefont
  {Matthiesen}}, \bibinfo {author} {\bibfnamefont {A.~N.}\ \bibnamefont
  {Vamivakas}}, \ and\ \bibinfo {author} {\bibfnamefont {M.}~\bibnamefont
  {Atat{\"u}re}},\ }\href@noop {} {\bibfield  {journal} {\bibinfo  {journal}
  {Physical review letters}\ }\textbf {\bibinfo {volume} {108}},\ \bibinfo
  {pages} {093602} (\bibinfo {year} {2012})}\BibitemShut {NoStop}%
\bibitem [{\citenamefont {Oulton}\ \emph
  {et~al.}(2002{\natexlab{b}})\citenamefont {Oulton}, \citenamefont {Finley},
  \citenamefont {Ashmore}, \citenamefont {Gregory}, \citenamefont {Mowbray},
  \citenamefont {Skolnick}, \citenamefont {Steer}, \citenamefont {Liew},
  \citenamefont {Migliorato},\ and\ \citenamefont {Cullis}}]{oulton2002}%
  \BibitemOpen
  \bibfield  {author} {\bibinfo {author} {\bibfnamefont {R.}~\bibnamefont
  {Oulton}}, \bibinfo {author} {\bibfnamefont {J.}~\bibnamefont {Finley}},
  \bibinfo {author} {\bibfnamefont {A.}~\bibnamefont {Ashmore}}, \bibinfo
  {author} {\bibfnamefont {I.}~\bibnamefont {Gregory}}, \bibinfo {author}
  {\bibfnamefont {D.}~\bibnamefont {Mowbray}}, \bibinfo {author} {\bibfnamefont
  {M.}~\bibnamefont {Skolnick}}, \bibinfo {author} {\bibfnamefont
  {M.}~\bibnamefont {Steer}}, \bibinfo {author} {\bibfnamefont {S.-L.}\
  \bibnamefont {Liew}}, \bibinfo {author} {\bibfnamefont {M.}~\bibnamefont
  {Migliorato}}, \ and\ \bibinfo {author} {\bibfnamefont {A.}~\bibnamefont
  {Cullis}},\ }\href@noop {} {\bibfield  {journal} {\bibinfo  {journal}
  {Physical Review B}\ }\textbf {\bibinfo {volume} {66}},\ \bibinfo {pages}
  {045313} (\bibinfo {year} {2002}{\natexlab{b}})}\BibitemShut {NoStop}%
\bibitem [{\citenamefont {Kuhlmann}\ \emph {et~al.}(2013)\citenamefont
  {Kuhlmann}, \citenamefont {Houel}, \citenamefont {Ludwig}, \citenamefont
  {Greuter}, \citenamefont {Reuter}, \citenamefont {Wieck}, \citenamefont
  {Poggio},\ and\ \citenamefont {Warburton}}]{kuhlmann2013}%
  \BibitemOpen
  \bibfield  {author} {\bibinfo {author} {\bibfnamefont {A.~V.}\ \bibnamefont
  {Kuhlmann}}, \bibinfo {author} {\bibfnamefont {J.}~\bibnamefont {Houel}},
  \bibinfo {author} {\bibfnamefont {A.}~\bibnamefont {Ludwig}}, \bibinfo
  {author} {\bibfnamefont {L.}~\bibnamefont {Greuter}}, \bibinfo {author}
  {\bibfnamefont {D.}~\bibnamefont {Reuter}}, \bibinfo {author} {\bibfnamefont
  {A.~D.}\ \bibnamefont {Wieck}}, \bibinfo {author} {\bibfnamefont
  {M.}~\bibnamefont {Poggio}}, \ and\ \bibinfo {author} {\bibfnamefont {R.~J.}\
  \bibnamefont {Warburton}},\ }\href@noop {} {\bibfield  {journal} {\bibinfo
  {journal} {Nature Physics}\ }\textbf {\bibinfo {volume} {9}},\ \bibinfo
  {pages} {570} (\bibinfo {year} {2013})}\BibitemShut {NoStop}%
\bibitem [{\citenamefont {M{\"u}ller}\ \emph
  {et~al.}(2012{\natexlab{c}})\citenamefont {M{\"u}ller}, \citenamefont
  {Bechtold}, \citenamefont {Ruppert}, \citenamefont {Zecherle}, \citenamefont
  {Reithmaier}, \citenamefont {Bichler}, \citenamefont {Krenner}, \citenamefont
  {Abstreiter}, \citenamefont {Holleitner}, \citenamefont {Villas-Boas} \emph
  {et~al.}}]{muller2012electrical}%
  \BibitemOpen
  \bibfield  {author} {\bibinfo {author} {\bibfnamefont {K.}~\bibnamefont
  {M{\"u}ller}}, \bibinfo {author} {\bibfnamefont {A.}~\bibnamefont
  {Bechtold}}, \bibinfo {author} {\bibfnamefont {C.}~\bibnamefont {Ruppert}},
  \bibinfo {author} {\bibfnamefont {M.}~\bibnamefont {Zecherle}}, \bibinfo
  {author} {\bibfnamefont {G.}~\bibnamefont {Reithmaier}}, \bibinfo {author}
  {\bibfnamefont {M.}~\bibnamefont {Bichler}}, \bibinfo {author} {\bibfnamefont
  {H.}~\bibnamefont {Krenner}}, \bibinfo {author} {\bibfnamefont
  {G.}~\bibnamefont {Abstreiter}}, \bibinfo {author} {\bibfnamefont
  {A.}~\bibnamefont {Holleitner}}, \bibinfo {author} {\bibfnamefont
  {J.}~\bibnamefont {Villas-Boas}},  \emph {et~al.},\ }\href@noop {} {\bibfield
   {journal} {\bibinfo  {journal} {Physical review letters}\ }\textbf {\bibinfo
  {volume} {108}},\ \bibinfo {pages} {197402} (\bibinfo {year}
  {2012}{\natexlab{c}})}\BibitemShut {NoStop}%
\bibitem [{\citenamefont {M{\"u}ller}\ \emph
  {et~al.}(2012{\natexlab{d}})\citenamefont {M{\"u}ller}, \citenamefont
  {Bechtold}, \citenamefont {Ruppert}, \citenamefont {Hautmann}, \citenamefont
  {Wildmann}, \citenamefont {Kaldewey}, \citenamefont {Bichler}, \citenamefont
  {Krenner}, \citenamefont {Abstreiter}, \citenamefont {Betz} \emph
  {et~al.}}]{muller2012high}%
  \BibitemOpen
  \bibfield  {author} {\bibinfo {author} {\bibfnamefont {K.}~\bibnamefont
  {M{\"u}ller}}, \bibinfo {author} {\bibfnamefont {A.}~\bibnamefont
  {Bechtold}}, \bibinfo {author} {\bibfnamefont {C.}~\bibnamefont {Ruppert}},
  \bibinfo {author} {\bibfnamefont {C.}~\bibnamefont {Hautmann}}, \bibinfo
  {author} {\bibfnamefont {J.}~\bibnamefont {Wildmann}}, \bibinfo {author}
  {\bibfnamefont {T.}~\bibnamefont {Kaldewey}}, \bibinfo {author}
  {\bibfnamefont {M.}~\bibnamefont {Bichler}}, \bibinfo {author} {\bibfnamefont
  {H.}~\bibnamefont {Krenner}}, \bibinfo {author} {\bibfnamefont
  {G.}~\bibnamefont {Abstreiter}}, \bibinfo {author} {\bibfnamefont
  {M.}~\bibnamefont {Betz}},  \emph {et~al.},\ }\href@noop {} {\bibfield
  {journal} {\bibinfo  {journal} {Physical Review B}\ }\textbf {\bibinfo
  {volume} {85}},\ \bibinfo {pages} {241306} (\bibinfo {year}
  {2012}{\natexlab{d}})}\BibitemShut {NoStop}%
\bibitem [{\citenamefont {Quilter}\ \emph {et~al.}(2013)\citenamefont
  {Quilter}, \citenamefont {Coles}, \citenamefont {Ramsay}, \citenamefont
  {Fox},\ and\ \citenamefont {Skolnick}}]{quilter2013enhanced}%
  \BibitemOpen
  \bibfield  {author} {\bibinfo {author} {\bibfnamefont {J.~H.}\ \bibnamefont
  {Quilter}}, \bibinfo {author} {\bibfnamefont {R.}~\bibnamefont {Coles}},
  \bibinfo {author} {\bibfnamefont {A.}~\bibnamefont {Ramsay}}, \bibinfo
  {author} {\bibfnamefont {A.}~\bibnamefont {Fox}}, \ and\ \bibinfo {author}
  {\bibfnamefont {M.}~\bibnamefont {Skolnick}},\ }\href@noop {} {\bibfield
  {journal} {\bibinfo  {journal} {Applied Physics Letters}\ }\textbf {\bibinfo
  {volume} {102}},\ \bibinfo {pages} {181108} (\bibinfo {year}
  {2013})}\BibitemShut {NoStop}%
\bibitem [{\citenamefont {M{\"u}ller}\ \emph
  {et~al.}(2013{\natexlab{c}})\citenamefont {M{\"u}ller}, \citenamefont
  {Kaldewey}, \citenamefont {Ripszam}, \citenamefont {Wildmann}, \citenamefont
  {Bechtold}, \citenamefont {Bichler}, \citenamefont {Koblm{\"u}ller},
  \citenamefont {Abstreiter},\ and\ \citenamefont {Finley}}]{muller2013all}%
  \BibitemOpen
  \bibfield  {author} {\bibinfo {author} {\bibfnamefont {K.}~\bibnamefont
  {M{\"u}ller}}, \bibinfo {author} {\bibfnamefont {T.}~\bibnamefont
  {Kaldewey}}, \bibinfo {author} {\bibfnamefont {R.}~\bibnamefont {Ripszam}},
  \bibinfo {author} {\bibfnamefont {J.}~\bibnamefont {Wildmann}}, \bibinfo
  {author} {\bibfnamefont {A.}~\bibnamefont {Bechtold}}, \bibinfo {author}
  {\bibfnamefont {M.}~\bibnamefont {Bichler}}, \bibinfo {author} {\bibfnamefont
  {G.}~\bibnamefont {Koblm{\"u}ller}}, \bibinfo {author} {\bibfnamefont
  {G.}~\bibnamefont {Abstreiter}}, \ and\ \bibinfo {author} {\bibfnamefont
  {J.}~\bibnamefont {Finley}},\ }\href@noop {} {\bibfield  {journal} {\bibinfo
  {journal} {Scientific reports}\ }\textbf {\bibinfo {volume} {3}} (\bibinfo
  {year} {2013}{\natexlab{c}})}\BibitemShut {NoStop}%
\end{thebibliography}%

\pagebreak

\section{Supplementary: Controlled tunneling induced dephasing of Rabi rotations for ultra-high fidelity hole spin initialization}


The supplementary material is organized as follows. In the first section, we present CW photocurrent absorption spectra of the QD, that allow to extract the electron tunneling times from the QD for high electric fields. The second section will briefly explain the rate equation model used to describe the population dynamics of the crystal ground state, the neutral exciton and the heavy hole state of the QD. In the third section, we will demonstrate ultrafast gating of the Schottky diode and in section four we use this gating technique to measure the fine struture precession of the neutral exciton. 

\section{CW photocurrent absorption measurements}

\begin{figure}[h]
\includegraphics[width=1\columnwidth]{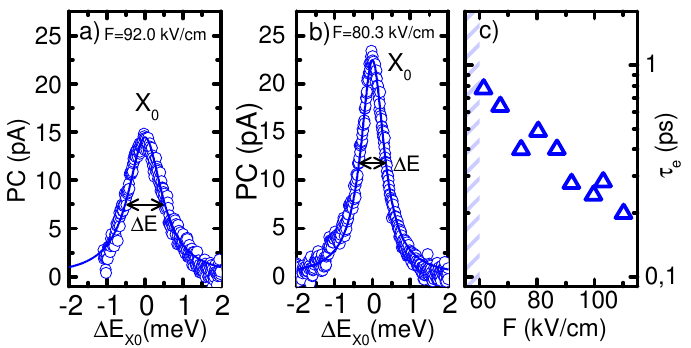}
\caption{\label{fig:Figure1s} CW-PC absorption spectra of the neutral exciton transition $X_{0}$ for electric fields of $F= 92.0 kV/cm$ in (a) and $F=80.3 kV/cm$ in (b). Due to ultrafast electron tunneling from the neutral exciton state $X_{0}$, the PC absorption spectrum experiences homegenous linewidth broadening. Lorentian fits of the absorption spectrum with homogenously broadened linewidth $\Delta E$ are presented in solid lines. In (c) the electron tunneling times calculated from the linewidth broadening as a function of electric field are presented.}
\end{figure}

In order to extract the tunneling times $\tau_{e}$ of the electron from the neutral exciton state $X_{0}$ at electric fields larger than $F= 50 kV/cm$, we recorded CW photocurrent (PC) absorption spectra of the neutral exciton transiton $cgs \rightarrow X_{0}$. For electric fields above $\sim 50 kV/cm$, electron tunneling occurs on timescales of less than $\tau_{e} < 5ps$ (see main section Figure 2), i.e. the electron tunneling time $\tau_{e}$ becomes approximately two orders of magnitude shorter than the neutral exciton radiative recombination time $\tau_{rec} \sim 400-800ps$ \cite{muller2007} and more than one order of magnitude shorter than the neutral exciton coherence time $\tau_{coh}$ \cite{muller2007, matthiesen2012}. Thus, the electron tunneling time $\tau_{e}$ can be treated as the main source of linewidth broadening \cite{oulton2002}. Consequently, in this regime where $\tau_{e}<<\tau_{rec},\tau_{coh}$, an observed homogenous linewidth broadening of the neutral exciton transition $X_{0}$ is directly linked to the reduced electron tunneling time $\tau_{e}$.   

In Figure \ref{fig:Figure1s}a and b, we present the measured CW-PC absorption spectra at electric fields of $F=80.3 kV/cm$ and $F=90.0 kV/cm$ for the neutral exciton transition $X_{0}$, respectively. Note, that the spectra have been corrected for a linear PC backround due to leakage current across the Schottky diode structure at high electric fields. While both spectra of the neutral exciton transition $X_{0}$ in Figure \ref{fig:Figure1s}a and b exhibit a clear Lorentian absorption line shape as indicated by the fit, the linewidth of the absorption peak in at an electric field $F=92.0 kV/cm$ (\ref{fig:Figure1s}) is broadened compared to the linewidth at an electric field of $F=82.3 kV/cm$ (\ref{fig:Figure1s}b). 

Assuming that the elctron tunneling is the main source of linewidth broadening \cite{oulton2002}, we calculate the electron tunneling times $\tau_{e}$ from the extracted linewidth $\Delta E$ via the Heisenberg uncertainty relation $\Delta E \cdot \tau_{e} = \frac{\hbar}{2}$. In Figure \ref{fig:Figure1s}c, we present the electric field dependence of the extracted electron tunneling times $\tau_{e}$. We neglected further sources of linewidth broadening e.g. electric noise \cite{kuhlmann2013} and power broadening \cite{muller2007} as they are much smaller than the homogenous broadening. This justification is consistent with the monotonic decrease of the tunneling time $\tau_{e}$ with increasing electric field $F$ and in good agreement with the WKB approximations and the pump-probe data presented in Figure 2 of the main section of the manuscript.

\section{Rate equation model for the charge occupation of the QD}

To quantitavely analyze the rise time and decay time of the PC ampltitude of the $X_{+}$ transition presented in Figure 1c of the main section, we use a rate equation model for the population of the QD. The QD states involved are the $cgs$, $X_{0}$ and $h^{+}$ as schematically presented in the main section of the manuscript. Since electron tunneling occurs on more than  three orders of magnitude faster timescales than hole tunneling $\tau_{e}<<\tau_{h}$,  we neglect the occupation of the QD with a single electron $e$ (that in principle could be created by hole tunneling $\tau_{h}$ from the neutral exciton $X_{0}$). The set of differential equation describing the population dynamics of the QD then reads \cite{muller2012electrical, muller2012high}:

\begin{widetext}
\begin{equation}
\frac{d}{dt}
\begin{pmatrix}
N_{X_{0}}(t)\\
N_{h^{+}}(t)\\
N_{cgs}(t)
\end{pmatrix}=
\begin{pmatrix}
	-(\Gamma_{e}+\Gamma_{rec}) & 0 & -\Gamma_{rec} \\
	\Gamma_{e} & -\Gamma_{h} & 0 \\
	\Gamma_{rec} & \Gamma_{h} & 0
\end{pmatrix}
\cdot
\begin{pmatrix}
	N_{X^{0}}(t)\\
N_{h^{+}}(t)\\
N_{cgs}(t)
\end{pmatrix}
\end{equation}
\end{widetext}

where the rates are related to the tunneling times of electrons and holes by $\Gamma_{e} = \frac{1}{ \tau_{e}}$ and $\Gamma_{h}=\frac{1}{\tau_{h}}$. Furthermore, we implented the radiative recombination of the neutral exciton $X_{0}$ with the rate $\Gamma_{rec}=800ps^{-1}$. For initial occupations of the $X_{0}$ and $cgs$ level of $N_{X_{0}}(t=0)=N_{0}$ and $N_{cgs}(t=0)=1-N_{0}$ created by the first excitation pulse, this set of differential equations can be solved analyically to obtain the following fitting functions for the PC-absorption amplitude \cite{muller2012electrical, muller2012high}:

\begin{widetext}
\begin{align}
I_{X^{0}\rightarrow 2X}(t)&=I_{0}+\Delta I_{0}\cdot exp(-(\Gamma_{e}+\Gamma_{rec})t) \label{eq:rateModelFitX0_2X}\\
I_{h^{+}\rightarrow X^{+}}(t)&=I_{0}+\Delta I_{0}\cdot\alpha\cdot[exp(-(\Gamma_{e}+\Gamma_{rec})t)-exp(-\Gamma_{h}t)] \label{eq:rateModelFith_Xp}\\
\begin{split}
I_{cgs\rightarrow X^{0}}(t)&=I_{0}-\Delta I_{0}\cdot[(\alpha+1)\cdot exp(-(\Gamma_{e}+\Gamma_{rec})t)-\alpha\cdot exp(-\Gamma_{h^{+}}t)]\\
&\quad-\Delta I_{0}\cdot exp(-(\Gamma_{e}+\Gamma_{rec})t)
\end{split} \label{eq:rateModelFitcgs_X0}
\end{align}
\end{widetext}

with $\alpha=\frac{\Gamma_{e}}{\Gamma_{h}-\Gamma_{e}-\Gamma_{rec}}$ and $\Delta I_{0} \propto N_{0}$. Note, that we added an offset $I_{0}$ to describe the experimentally observed backround in the PC due to leakage currents across the Schottky diode structure. By fitting the temporal decay of the $X_{+}$ transition as presented in Figure 1d in the main section, we extract the hole tunneling time $\tau_{h}$ for each electric field $F$.

\section{Effectiv Schottky diode switching time}

\begin{figure}[h]
\includegraphics[width=1\columnwidth]{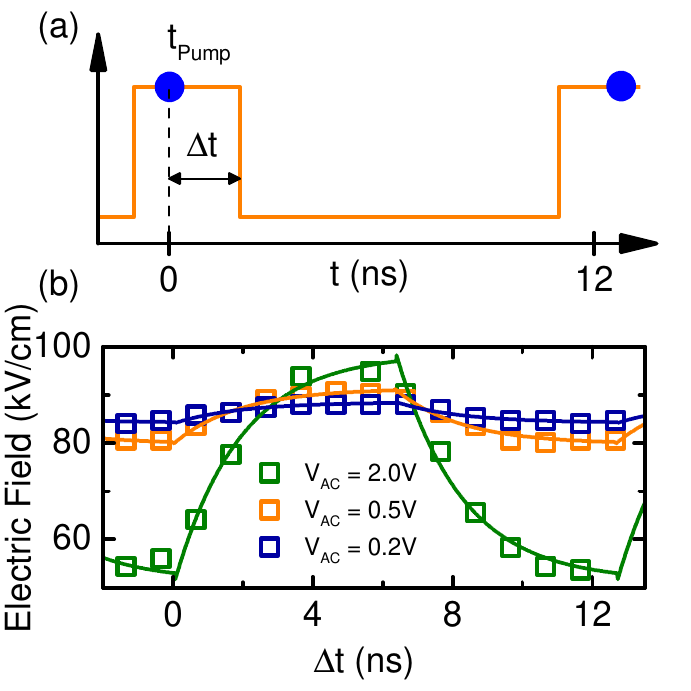}
\caption{\label{fig:Figure3s} (a) Switching scheme (b) Electric field modulation extracted from the Stark shift of the neutral exciton $X_{0}$.}
\end{figure}

In order to access the effective switching time of the Schottky diode, we modulate the electric field across the QD layer using a pulse pattern generator which creates a square wave bias modulation, that is added to the DC bias applied to the Schottky diode. The modulation bias is synchronized to the pulsed laser excitation. As illustrated in figure \ref{fig:Figure3s}a, we extract the diode repsonse by measuring the resonance energy of the neutral exciton transition $cgs \rightarrow X_{0}$ as a function of the time delay $\Delta t$ between a single excitation pulse (pump) and the square wave modulation from bis the pulse pattern generator.  

The electric field at the QD is calculated from the resonance energy of the neutral exciton $X_{0}$ using the Quantum confined Stark effect \cite{Krenner2005}. In figure \ref{fig:Figure3s}b, we present the extracted electric fields for square wave modulation of  $V_{AC} = 2.0V, 0.5V, 0.2V$ on top of a DC bias of $V_{DC}=-1.5V$. The data is fitted using the response function of an RC-circuit. Due to imperfect impedance matching of the electronic circuit, only $1/2$ of the modulation amplitude of the square wave bias is transfered to the QD. However, the electronic circuit displays an RC constant of $1.87ns$; well below the hole tunneling time $\tau_{h}$ in the corresponding electric field region. Thus, switching the diode structure to e.g. electric field regions with increased hole storage times after the hole spin initialization or applying a reset pulse that empties the QD from charge carriers during every cycle \cite{quilter2013enhanced} is easily feasible. 

Note here, that although in general long hole storage times $\tau_{h}$ are desirable, we tailored the Al-concentration of the hole tunnelling barrier to be $10 \%$ for a device with PC read out. Thus for electric fields $F > 50 kV/cm$ in Figure 2a of the main section of the article, the electron tunneling time $\tau_{e}$ is on the timescale of the excitation laser pulse length $\tau_{pulse}$, while the hole storage time $\tau_{h}$ matches the repetition time of the excitation laser pulse of 12.6ns. Since the PC read-out signal is proportional to the number of optically generated charge carriers per time, a hole storage time $\tau_{h}$ longer than the laser pulse repetition rate diminishes the PC signal as not every pulse would optically generate new charge carriers.

\section{Measurements of the fine structure precession of the neutral exciton $X_{0}$}

In order to determine the finestructure precession period $\tau_{fss}$ limiting the hole spin initialization fidelity at zero magnetic field, we perform pump-probe measurements at an electric field $F=30.9 kV/cm$. At this electric field the hole tunnelling time is much longer tham the repetition rate of the laser, suppressing PC read out. We add a voltage modulation locked to the repetition rate of the excitation laser with an amplitude of $1.5V$ to the DC voltage across the Schottky diode structure \cite{quilter2013enhanced}. The electric field across the diode is increased during the seond half of the $12.6ns$ cycle to $F>90 kV/cm$ such that remaining holes are "kicked out", after the exciton dynamics have been probed with a pump and probe pulse during the first half of the $12.6ns$ cycle at an electric field of $F=30.9 kV/cm$. Note, that at $F=90.9 kV/cm$ the hole tunnelling time $\tau_{h}$ exceeds the laser pulse repetition rate of $79 MHz$.

\begin{figure}[t]
\includegraphics[width=1\columnwidth]{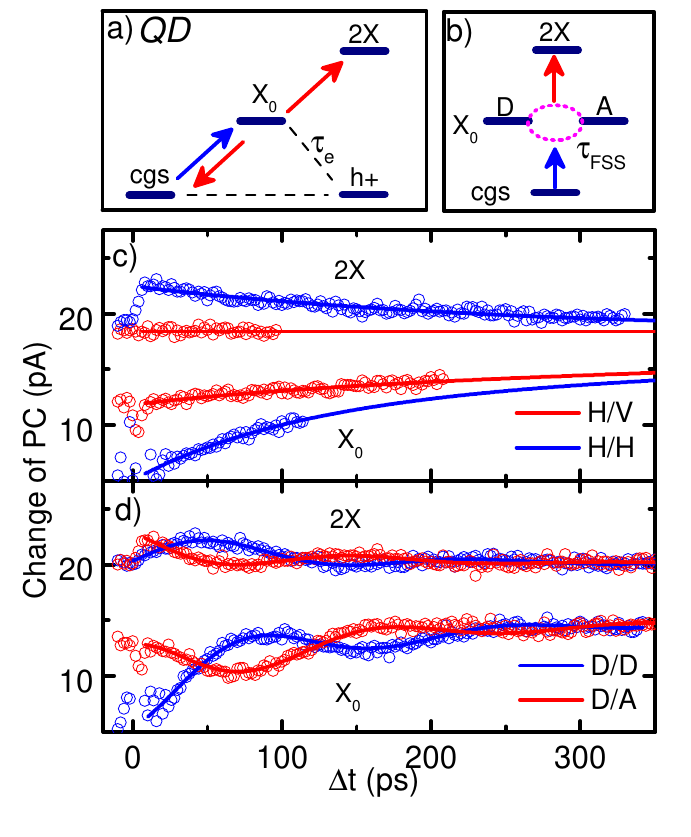}
\caption{\label{fig:Figure2s} (a) and (b) Schematic representation of the excitation scheme with two pulses indicated in blue and red for a QD initially in the $cgs$.While the first pulse (blue) transfers the population from the $cgs$ to the $X_{0}$, the second pulse (red) probes the population of the QD by stimulated emssion to the $cgs$ or biexciton generation 2X. For a diagonally polarized pulse $D$ the spin of the generated exciton $X_{0}$ precesses with the finestructure period $\tau_{fss}$ (c) and (d) Time evolution of the exciton $X_{0}$ and biexciton $2X$ PC transition amplitude under co- and cross-linearly polarized two pulse excitation for pump-probe delays $\Delta t$. The antiphased oscillations in (d) reflect the finestructure precession of the $X_{0}$ spin states, that were generated in a coherent superposition D of the spin eigenstates H and V addressed in (c).}
\end{figure}

The scheme used to determine the finestructure precession period $\tau_{fss}$ is illustrated in Figure \ref{fig:Figure2s}a. We excite the QD, initially in the $cgs$, with two laser pulses indicated as blue and red arrows in Figure \ref{fig:Figure2s}a delayed by $\Delta t$ with respect to each other. Note, that here $\Delta t$ is shorter than the electron tunneling time $\tau_{e}$. The first pump pulse (blue) transfers the population of the QD from the $cgs$ to the neutral exciton $X_{0}$. While the second laser pulse consequently probes the occupation of the QD. If the second laser pulse is energetically in resonance with the neutral exciton, stimulated emission can occur if the polarization of the laser pulse matches the exciton spin state. The probe pulse is energetically tuned into resonance with the red detuned biexciton transition $X_{0} \rightarrow 2X$, the population of the QD will similiarly be driven to the biexciton state $2X$ depending on the polarization of the probe pulse \cite{muller2013all}.

In Figure \ref{fig:Figure2s}c and d, we present the temporal evolution of the amplitudes of the neutral exciton transition $cgs \rightarrow X_{0}$ and biexciton transition $X_{0} \rightarrow 2X$ (plotted as the change of PC $\Delta PC = PC_{pump-probe} - PC_{pump} $). While for excitation with two co-linearly polarized pulses $H/H$ in Figure \ref{fig:Figure2s}c, we observe an exponentially decaying amplitude of the $X_{0}$ and $2X$ transition, for cross-linear polarized excitation with $H/V$, the amplitude of both transitions is surpressed. 

The surppression of the transition amplitudes can be understood as follows: The pump pulse with  a polarization $H$ creates an exciton in one of the eigenstates of finestructure split energy level $X_{0}$ that reads $\frac{1}{\sqrt{2}}(\uparrow \Downarrow + \Uparrow \downarrow)$. Here, $\downarrow (\uparrow)$ denotes the electron spin and $\Uparrow (\Downarrow)$ the hole spin. In order to generate a biexciton in the only allowed Pauli spin configuration $\downarrow \uparrow \Downarrow \Uparrow$, the pump pulse has to be polarized along $H$. Thus, the absorption of a $V$ polarized probe pulse is blocked due to the Pauli exclusion principle and the amplitude of the $X_{0} \rightarrow 2X$ transition in Figure \ref{fig:Figure2s}c is surppressed \cite{muller2013all}. 

In analogy to  the optical selection rules for the $X_{0} \rightarrow 2X$ transition, the second laser pulse can only transfer population from the $X_{0}$ to the crystal ground state $cgs$ via stimulated emission, if it has the same polarization as the first pulse. Consequently, the amplitude of the bleaching at the $X_{0}$ transition in Figure \ref{fig:Figure2s}c, is reduced for cross-linearly polarized excitation with $H/V$. 

In Figure \ref{fig:Figure2s}d, we present the time dependence of the neutral exciton transition amplitude $X_{0}$ and biexciton transition amplitude $2X$ excited with diagonally polarized light $D=\frac{1}{\sqrt{2}}(H+V)$ and $A=\frac{1}{\sqrt{2}}(H-V)$. Additionally to the exponential decay of the amplitudes oberserved in Figure \ref{fig:Figure2s}c, we observe an antiphased oscillation for excitations with $D/D$ and $D/A$. As we excite the $cgs \rightarrow X_{0}$ transition with a D-polarized pump pulse, the pump pulse creates a coherent superposition of the neutral excition spin eigentstates $H=\frac{1}{\sqrt{2}}(\uparrow \Downarrow + \Uparrow \downarrow)$ and $V=\frac{1}{\sqrt{2}}(\downarrow \Uparrow - \uparrow \Downarrow)$ namely $D=\frac{1-i}{\sqrt{2}}(\uparrow \Downarrow +i \downarrow \Uparrow)$. Due to the electron hole exchange interaction, the exciton spin superposition state will start to precess from $D=\frac{1-i}{\sqrt{2}}(\uparrow \Downarrow +i \downarrow \Uparrow) \rightarrow L=\uparrow\Downarrow \rightarrow A=\frac{1+i}{\sqrt{2}}(\uparrow \Downarrow - i \downarrow \Uparrow) \rightarrow R=\Uparrow\downarrow \rightarrow D=\frac{1-i}{\sqrt{2}}(\uparrow \Downarrow +i \downarrow \Uparrow)$ as illustrated in Figure \ref{fig:Figure2s}b \cite{muller2013all}. However, as the optical selection rules only allow the generation of the biexciton $2X$ (stimulated emission of the $X_{0}$) for co-linearly polarized pump-probe pulses $D/A$ ($D/D$), the finestructure precession of the exciton spin states translates into an oscillation of the corresponding PC amplitude. Thus, by probing the biexciton transition $X_{0} \rightarrow 2X$ with polarizations $D$ and $A$ (or the stimulated emission the transition $X_{0} \rightarrow cgs$), we observe an antiphased oscillation in Figure \ref{fig:Figure2s}d. From the oscillation period, we extract a finestructure precession time of $\tau_{fss}= 175 \pm 3ps$ that corresponds to a finestucture energy splitting of $\hbar \delta_{fss} = 23.6 \pm 0.4\mu eV$.

\end{document}